\documentclass[a4paper, 11pt]{article}

\usepackage[german, english]{babel}
\usepackage{authblk}
\usepackage{setspace}
\usepackage{array}
\usepackage{tabularx}
\usepackage{graphicx}
\usepackage{amsmath}
\usepackage[left=2.0cm,right=2.0cm,top=2.0cm,bottom=2.0cm]{geometry}
\usepackage{float}
\usepackage[authoryear,round]{natbib}
\usepackage{url}
\usepackage{doi}
\usepackage{xcolor}
\usepackage{makecell}

\newcommand\authorcontribution{\endlist\egroup}
\newcommand\competinginterests{\endlist\egroup}
\newcommand\acknowledgments{\endlist\egroup}

\begin{document}

\doublespacing

\title{Advancing global water cycle science with the upcoming MAGIC satellite mission through data assimilation}
\author[1]{Helena Gerdener}
\author[1]{Yorck Ewerdwalbesloh}
\author[1]{Anne Springer}
\author[1]{Annika Nitschke}
\author[1]{Jürgen Kusche}

\affil[1]{Institute of Geodesy and Geoinformation, University of Bonn, Bonn, Germany}
\affil[ ]{Corresponding author: Yorck Ewerdwalbesloh, ewerdwalbesloh@geod.uni-bonn.de}

\date{ }

\maketitle

\begin{abstract}
The GRACE and GRACE-FO missions provide unique global measurements of terrestrial water storage anomalies (TWSA) from space since 2003. However, in order to spatially downscale the GRACE/-FO resolution of 300 km further, and to vertically disaggregate TWSA into soil moisture, groundwater and other contributions, the TWSA observations were assimilated into hydrological and land surface models in the past. This, at the same time, also improves the realism of the models. 
With the upcoming MAGIC (Mass Change and Geosciences International Constellation) constellation consisting of two satellite pairs (NASA/DLR GRACE-C and ESA NGGM), significant advancements are expected for the monitoring of water mass redistribution processes within the Earth system, with improved temporal and spatial resolution, reduced latency, and with lower noise levels. In this study, we reviewed the capabilities and limitations of existing GRACE/-FO data assimilation systems. We then investigated the benefit of including data from the upcoming MAGIC constellation into existing assimilation frameworks to improve not only data sets for water and carbon cycle science but also in view of operational applications, within a closed-loop simulation. For example, we extended a global 17 years GRACE/-FO assimilation era to a long time series of 33 years with synthetic MAGIC data assimilated (which will be reached in mid 2030s) and show reductions in groundwater trends errors and errors in identifying hydrological extreme events. We further showed for Europe that both storage and river runoff simulations benefit from more frequent assimilation, which the integration of MAGIC data will enable due to the foreseen 5-daily (instead of monthly) TWSA product, and we found this independent of whether we used a univariate TWSA or a multivariate TWSA and soil moisture assimilation system. Finally, we demonstrated in a synthetic experiment for the Iberian Peninsula and for Eastern Europe that MAGIC TWSA data assimilation also improved simulated carbon cycle variables via coupling in the land surface model, especially in water-limited regions.
\end{abstract}

\section*{Keywords}
assimilation, MAGIC, terrestrial water storage anomalies, hydrological models, land surface models, GRACE/-FO

\newpage

\section{Introduction}
The GRACE (Gravity Recovery And Climate Experiment) and GRACE-FO (GRACE Follow-On) satellite gravity missions have provided unprecedented data to improve our understanding of the Earth system. GRACE and GRACE-FO (hereafter GRACE/-FO) enabled the comprehensive monitoring of ice sheet and glacier mass balances, of large-scale land water deficit or surplus related to meteorological extremes and climate events like the El Niño Southern Oscillation, of the barysteric contribution to sea level rise, and of anthropogenic groundwater abstraction \citep{rodell2023}. GRACE-/FO have transformed our view on the water cycle by uncovering long-term deviations of terrestrial water storage \citep{Rodell2018, chandanpurkar_unprecedented_2025}. As a result, total water storage (TWS) has been recently adopted by GCOS as an essential climate variable \citep{bojinski_concept_2014}. This recognizes its role in critically contributing to the characterization of Earth's climate, and places demands on documentation and transparency of its assembly from gravity observations, regular updates, and the provision of multiple data sets for the same variable.

The primary GRACE/-FO observable, temporal TWS anomalies (TWSA), has been used in studies of anthropogenic water use quantification \citep{scanlon_effects_2021, abdelmohsen_buffering_2022}, the linkages between deep groundwater, soil moisture, vegetation growth and evapotranspiration \citep{cook_efficacy_2021, gerdener2022}, climate variability and extremes \citep{humphrey_assessing_2016, giardina_large_2025}, drought monitoring and risk \citep{houborg_drought_2012, zhao_global_2017, gerdener_framework_2020}, flooding potential and inundated water volume \citep{reager_river_2014, han_grace_2021}, and hydrology and climate model evaluation \citep{scanlon2018global, jensen_long-term_2019, jensen_observations_2024}. Beyond hydrology and hydrometeorology, TWSA observations provide essential contributions in e.g. ocean and sea level science, geodesy, geodynamics, and Earth rotation studies \citep[review in, e.g.,][]{chen2022applications}.

With the GRACE/-FO missions, tiny intersatellite orbital variations of the spacecraft are measured with extremely high precision, and inverted to monthly gravity field maps and further to monthly TWSA maps. However, the spatial resolution of these maps is only about 300 km currently, due to instrument noise, the downward continuation from satellite altitude, and the presence of fast (i.e. timescales below one month) mass transports in ocean, atmosphere and hydrosphere that cannot be sufficiently corrected for and create aliasing effects. Moreover, due to its measurement principle, GRACE/-FO sense only the variability of the vertical total aggregate of all water storages. This means that for isolating groundwater or surface water volume changes, or root-zone soil moisture variability from GRACE-/FO observations, further analysis techniques need to be employed. Some studies derive groundwater storage changes as a residual in the terrestrial water budget \citep[e.g.,][]{getirana2025}, but this would require that we have unbiased and physically consistent estimates for soil moisture, surface water storage etc. from complementary observations and models.

In order to circumvent the coarse native resolution of GRACE/-FO data, which is also inherent in so-called mascon solutions, a number of studies apply machine learning, deep-learning or statistical downscaling approaches to the TWSA grids, typically relying on high-resolution precipitation and other climate predictor data \citep[e.g.,][]{li2026reproducing}. However, these studies meet challenges in fully representing random and systematic uncertainties, and it is often difficult to assess the systematic influence of predictor data, which renders their applicability for model evaluation challenging.

Therefore, an increasing number of studies developed formal data assimilation (DA) approaches for the integration of the GRACE/-FO TWSA maps into hydrological models or land surface models on a regional or a global scale, see e.g. the review in \cite{springer_review_2026}. DA does not only serve for partitioning TWSA to model storages in physically consistent ways, but also for spatially downscaling the coarse GRACE/-FO data to the finer model resolution. From a modeling perspective, DA improves the realism of model variables and mitigates the effect of forcing data and model structural errors. Data assimilation approaches also bridge several 1- or 2-month GRACE data gaps and the 11-month gap between the GRACE and the GRACE-FO missions dynamically consistent with the underlying hydrological models, however, these periods may require additional scrutiny when TWSA DA maps are used in hydrological applications.

The NASA/DLR GRACE Continuity (GRACE-C) mission, with orbital geometry and instruments nearly identical to its predecessors GRACE/-FO, is foreseen for launch in 2029 with a design lifetime until at least 2034. GRACE-C will thus extend the TWSA record with a potential overlap with GRACE-FO, which would potentially allow testing data sets of higher temporal resolution.
Further, ESA’s Next Generation gravity Mission (NGGM) is currently in phase B development and foreseen for launch in 2032, with the same measurement principle in an inclined orbit of lower altitude, and thus higher sensitivity compared to GRACE-C in the region bounded by latitudes $\pm 70^{\circ}$. Combining measurements from the two missions, as proposed by the ESA/NASA MAss Change and Geosciences International Constellation (MAGIC) Working Group, is expected to enable substantial improvements in spatial and temporal resolution and data latency \citep{daras2024mass}. While these new TWSA data are expected to be exploited in DA frameworks, it remains unclear whether and to what extent their assimilation translates into improvements in level-4 products relevant to water cycle research. This includes, for example, high-resolution estimates of root-zone soil moisture and groundwater storage, drought monitoring, and the representation of terrestrial biosphere responses to variations in precipitation and water availability.

In this contribution, we evaluate the impact of the upcoming MAGIC data record in typical assimilation applications at global, continental and regional scale. We provide a review of existing global, continental and regional DA systems and their limitations to insert the potential of MAGIC DA into context of existing DA systems. We then select a few key DA applications that we suggest will be relevant not only to water cycle science but also to operational applications under MAGIC: (1) Mid of the next decade, the TWSA record will be $>$30 year and enable us to look at significantly improve estimates of long-term average (i.e. climatological) trends and annual amplitudes in TWSA, groundwater, soil moisture, and surface water storage loss and gain. We quantify the benefit of having continued observations with MAGIC with the global data assimilation here, in addition to the GRACE/-FO record. We believe this showcases the representative uncertainty of upcoming climate-long reference data sets for water storages. We show that this will also improve drought monitoring applications, by providing a better estimate of the statistical distribution of water deficit.
(2) MAGIC will provide operational TWSA solutions with 5-day resolution. We quantify how assimilating these short-term solutions will improve estimates of TWSA trends and variability representation for major European river catchments in a high-resolution (12.5 km) continental-system DA framework as compared to monthly solutions. We find that the improved temporal resolution with MAGIC DA leads to an improved partitioning of TWSA, particularly in terms of root-zone soil moisture, and also benefits the simulation of basin discharge. This is also investigated in a scenario of multivariate DA where we, in addition to TWSA, assimilate surface soil moisture. (3) Finally, we study the impact of the assimilation of 5-daily MAGIC TWSA data on the representation of carbon cycle related variables such as leaf area index (LAI) and gross primary production (GPP) in regional model runs with activated biogeochemical cycling. We hypothesize here that the improved skills in the vertical soil moisture partitioning will aid in representing carbon turnover and below- and aboveground biomass production in the model, and thus may also pave the way for exploring synergies of TWSA and vegetation remote sensing.


\section{Materials and methods}


\subsection{Simulation setup}
\label{sec:simulator}
\begin{figure}
\centerline{\includegraphics[width=0.98\textwidth]{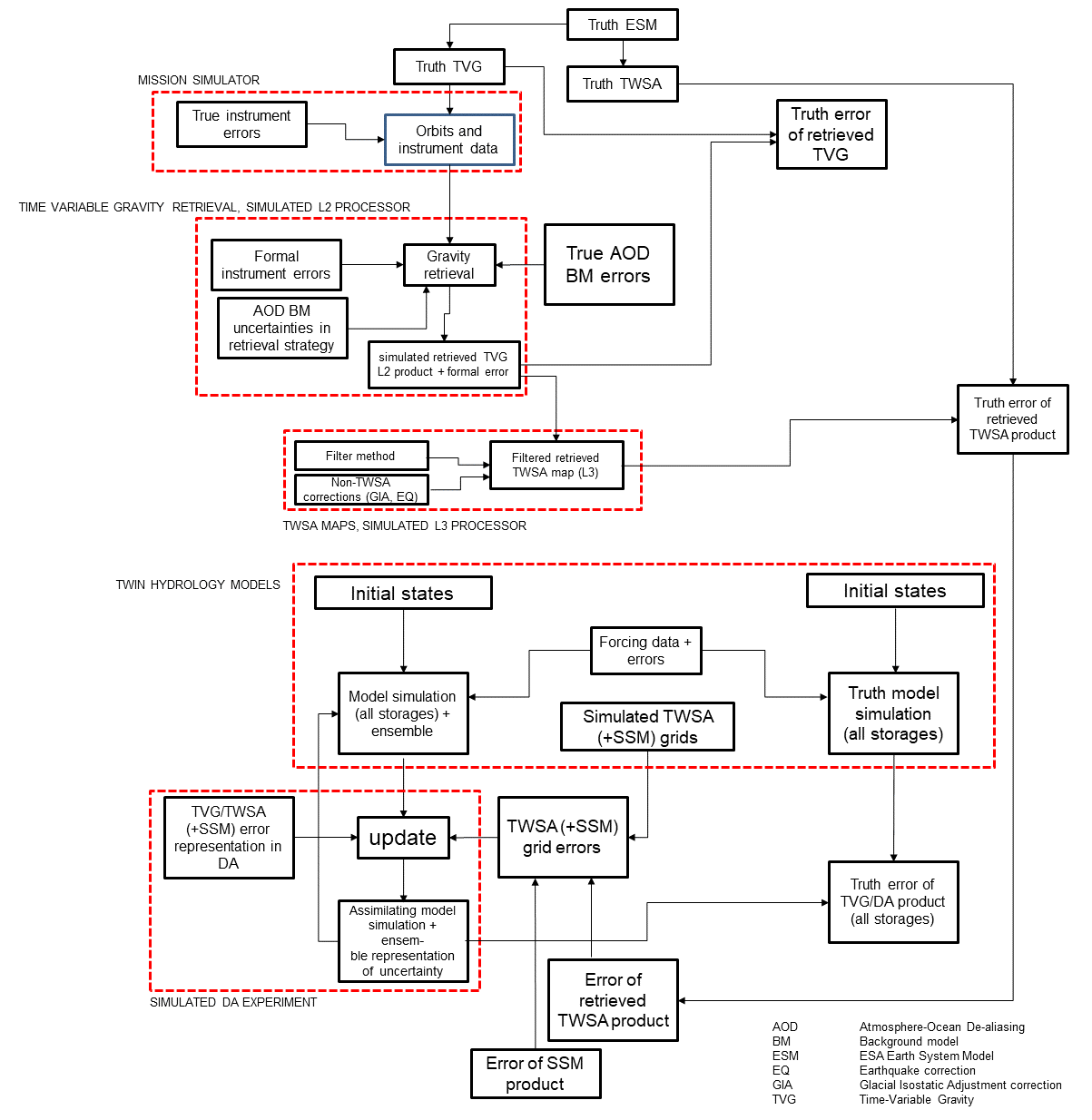}}
\caption{Simulation logic used in this study. Top part represents the closed-loop SING satellite data product simulator for deriving GRACE-C-like and MAGIC simulations, bottom part represents the twin hydrological models and DA OSSE experiment. SSM (Surface Soil Moisture) refers to simulations with additional satellite soil moisture assimilation.}
\label{fig:simulations}
\end{figure}

In the framework of the ESA \emph{Studying the Impact of the NGGM and MAGIC Gravity Missions (SING)} study, the added value of NGGM and MAGIC has been investigated for scientific applications and operational services in hydrology, ocean sciences, glaciology, climate sciences, solid earth sciences, and geodesy beyond GRACE-FO
using a closed-loop simulator with a comprehensive description of instrumental and dealiasing errors \citep{pfeffer_impact_2026}. 
This simulator (top part of Fig. 
\ref{fig:simulations}) 
provides synthetic observations for level-1 data ("orbits and instrument data"), and generates realistic outcomes of the level-2 ("TVG L2") and level-3 ("TWSA maps L3") data retrieval \citep{schlaak2026simulation}, including post-processed TWSA maps for GRACE-C-like and NGGM scenarios consistently during the overlap period. These synthetic data contain simulated errors from the GRACE-C-like and NGGM instrument noise, incomplete background models, and imperfect retrieval and postprocessing strategies \citep{schlaak_resolving_2025}. We here focus on using simulations from the MAGIC constellation, i.e. GRACE-C-like combined with NGGM, for presenting its potential for use in assimilation systems but we will also include GRACE-C-like scenarios for comparison.

In the end-to-end SING simulations, the gravity retrieval is implemented via the "acceleration difference approach" \citep{Murbock.2014,Schlaak.2023}. Within a least square adjustment, weekly and monthly sets of spherical harmonic coefficients are then retrieved by combining normal equations for the two missions. The truth gravitational fields are derived from the ESA Earth System Model (ESM), which is a time-variable mass transport model with 6-hour resolution \citep{dobslaw2015}. As in the GRACE-FO processing, atmosphere-ocean dealiasing errors are simulated by reducing observations with a perturbed correction model \citep{dobslaw2016}, taking de-aliasing errors \citep{shihora2024} and ocean tide model errors (model differences between EOT11a \citep{Savcenko.2012} and GOT4.7 \citep{Ray.2008} into account. While GRACE-C is assumed as flying at about 500 km altitude with an inclination of 89°, the NGGM orbit is simulated at about 397 km and with 70° inclination. 

Comparing truth TWSA from the ESA ESM model and the retrieved TWSA maps allows then to arrive at an empirical realization of truth TWSA errors along with an estimate of the MAGIC L2 and/or L3 variance-covariance matrix. More, specifically, the variance-covariance matrices (VCMs) for the GRACE-C-like and MAGIC simulations are determined in an empirical approach by using Monte Carlo simulations, i.e. 20.000 simulations for the weekly and 40.000 simulations for the monthly solution. In this approach, the VCMs are computed from VADER filtered \citep{horvath2018} residuals, i.e. that the VADER filtered MAGIC simulations are compared against the unfiltered defined truth, which is the ESA-ESM2.0 \citep{dobslaw2015}. This means that the empirical VCMs contain commission errors and errors from signal damping from the filtering. 

In this study, we will analyze the effect of MAGIC assimilation on storage and fluxes output in an Observing System Simulation Experiment (OSSE, bottom part of Fig.~\ref{fig:simulations}), while preserving the tuning established during the GRACE/GRACE-FO assimilation era to ensure consistency. We have four different DA systems at the Institute of Geodesy and Geoinformation, University of Bonn that assimilate GRACE/-FO TWSA into models (Sec~\ref{sec:DA_settings}), a global, two continental, and a regional DA system. First, the global WaterGAP assimilation system \citep{gerdener2023global,gerdener2024} with monthly water storage outputs at a spatial resolution of about 50 km. Second and third, the continental CLM3.5 and eCLM assimilation systems \citep{springer_evidence_2019, ewerdwalbesloh_grace_2026}, which focus on Europe and operate at a finer spatial resolution of about 12.5 km. Fourth, the regional eCLM assimilation System focusing on Europe, which operates at a fine spatial resolution of 3 km. All four systems are prepared with initial states and external forcing data to enable water storage simulation. In addition, the continental and regional eCLM DAs can additionally assimilate soil moisture, i.e. they are also multivariate systems.
    
In these real-world DA systems we exchange the observations now by a synthetic truth from one of the four models equipped with the GRACE-C-like or MAGIC mission scenario errors (cmp. TWSA grid errors). As usual during assimilation, the predicted model simulation (cmp. model simulation + ensemble) is then updated each time interval by weighting model errors with errors from the GRACE-C-like or MAGIC simulation. Thus, the simulated VCMs from the mission scenarios are used twice in our assimilation: 1) to mimic the observational error and 2) to perturb a hydrological truth (Sec.~\ref{sec:truth}).

It is important to understand that the hydrological truth in our OSSE refers to a model which can, unlike the hydrology model within ESA ESM \citep{dobslaw2015}, be run in DA mode. Next to this truth model, we run a second model with similar model structure which differs from the truth by random and systematic effects, and which we seek to move closer to the truth by assimilating simulated data. Finally, the similar model structures of the truth and assimilation models allow for a direct comparison not only of TWSA errors but also of all simulated model storages and fluxes. This would not be possible if the ESA ESM hydrology model was used as the concepts of e.g. soil moisture layers from the truth and from the DA would differ.

We expect the smaller the gravity simulation error, the closer the DA output will be towards the simulated truth. With these OSSEs we intend to analyze the effect of including MAGIC simulations in the assimilation frameworks on storage and fluxes outputs in comparison to GRACE-C-like assimilation scenarios or existing GRACE/-FO assimilation frameworks. The details of used models, deriving the simulated truth, equipping it with errors to mimic synthetic observations, and assimilation setup are described subsequently. 


\subsection{Hydrological and land surface models}
\label{sec:models}
The synthetic TWSA are assimilated into three hydrological and land surface models covering different spatial scales and levels of process complexity: the global hydrological model WaterGAP and two generations of the Community Land Model (CLM), CLM3.5 and the CLM5-based eCLM. WaterGAP is used for global experiments, whereas CLM3.5 and eCLM are applied over Europe at higher spatial and temporal resolution. This model hierarchy allows the potential benefits of the future satellite missions to be assessed from global-scale water storage estimation to the representation of individual hydrological processes at continental and regional scales. WaterGAP additionally accounts for anthropogenic water use, whereas the CLM models provide a more detailed representation of vertical soil water redistribution. The eCLM applications extend the analysis in two complementary directions: in a continental-scale setup, near-surface soil water content (SWC) can be assimilated in addition to TWSA within a multivariate data assimilation framework, whereas the regional setup simulates vegetation and carbon dynamics to investigate how TWSA DA affects them through their interactions with the terrestrial water cycle.

The Water--Global Assessment and Prognosis (WaterGAP) model is a global hydrological model that simulates daily water fluxes and storages over all continental land areas except Antarctica \citep{muellerschmied2021,muller2024}. It simulates water storage in 10 layers for soil, canopy, snow, lakes (global and local), wetlands (global and local), rivers, reservoirs, and groundwater. The model operates at a spatial resolution of 0.5°, corresponding to approximately 50 km, and comprises about 67.420 active land grid cells. WaterGAP combines the WaterGAP Global Hydrology Model (WGHM), which represents vertical and lateral water balances, with sector-specific water-use models and a linking module that assigns withdrawals and consumptive use to groundwater and surface water resources. Water use is represented for irrigation, livestock, domestic supply, manufacturing, and thermal power generation. A detailed description of WaterGAP is provided in \cite{muller2024} and its use in DA in \cite{gerdener2024}.

The Community Land Model version 3.5 (CLM3.5) is a land surface model that simulates terrestrial water and energy fluxes and storages \citep{oleson_improvements_2008}. It is applied over a European domain based on the EUR-11 grid \citep{jacob_regional_2020} at a spatial resolution of approximately 12.5 km. Land-surface heterogeneity is represented through a sub-grid description of different surface and vegetation types. Water storage comprises canopy water, snow represented by up to five layers, liquid water and ice in ten soil layers extending to a depth of approximately 3 m, and water stored in an unconfined aquifer. CLM3.5 represents interception, snow accumulation and melt, infiltration, vertical soil-water redistribution, surface and subsurface runoff, groundwater recharge, and drainage. The European CLM3.5 setup, including the atmospheric forcing and static input data, is described in detail in \cite{springer_water_2019}.

The encore Community Land Model (eCLM) is a fork of CLM5 with an adapted build infrastructure, while retaining the scientific core of the original model \citep{lawrence_community_2019,poll_new_2025}. CLM5 is a later generation of the Community Land Model that retains the general process-based representation of terrestrial water and energy fluxes such as CLM3.5, while incorporating updated representations of individual land-surface processes \citep{lawrence_community_2019}. Compared with CLM3.5, the fixed 3 m soil column and unconfined aquifer representation are replaced by up to 20 soil layers with a spatially variable depth determined by bedrock, reaching a maximum of 8.03 m. Snow processes are represented in greater detail using up to 12 snow layers, and an additional surface water storage is included. The plant hydraulic stress scheme \citep[PHS;][]{kennedy2019} simulates water transport from the soil via the plant to the leaves. When soil water becomes limiting, leaf water potential declines, which in turn reduces stomatal conductance and thereby lowering transpiration and photosynthetic activity. 

\begin{figure}[h!]
\begin{center}
\includegraphics[width=12cm]{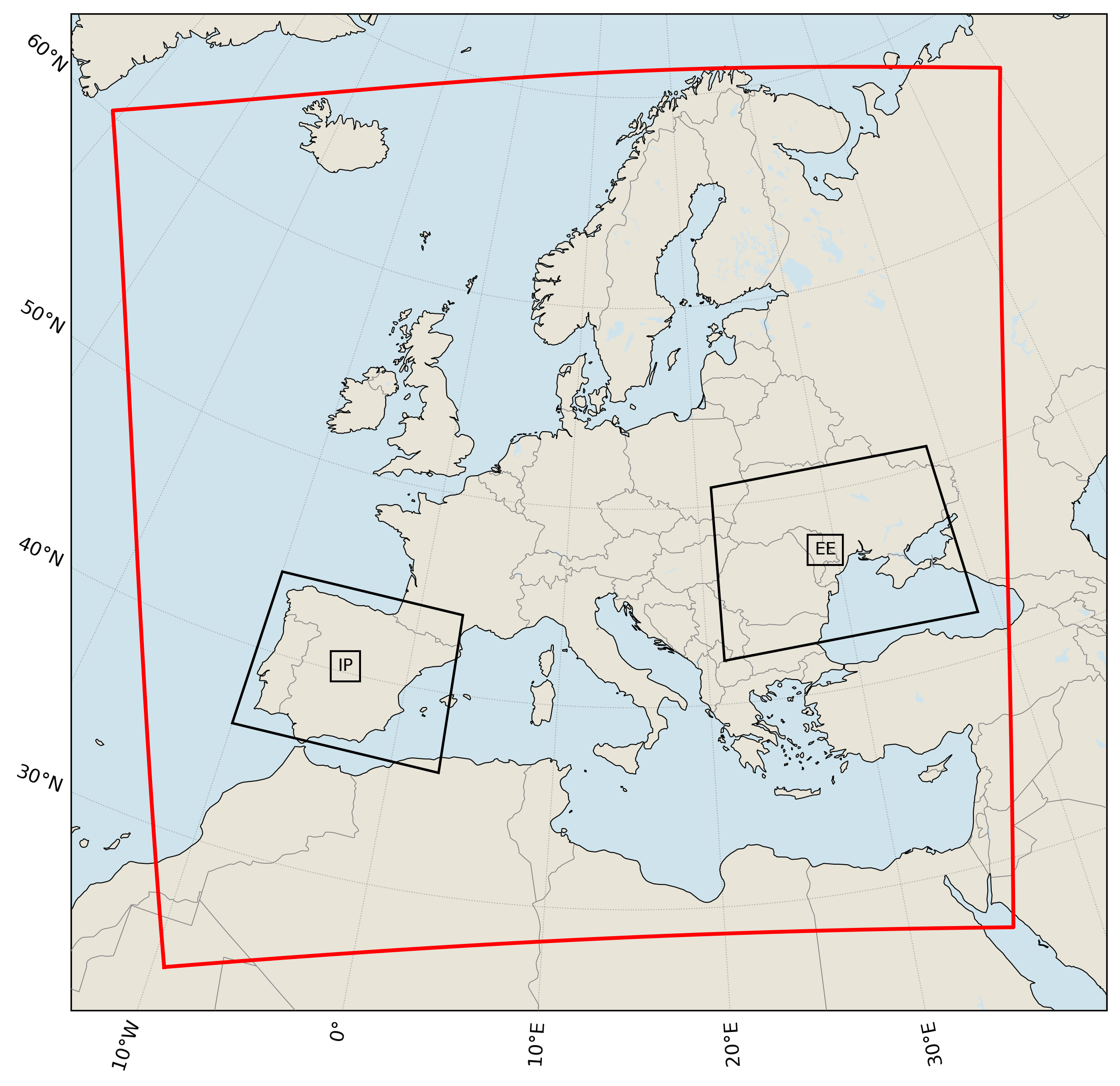}
\end{center}
\caption{Study domains used for the CLM3.5 and eCLM DA simulations. The red boundary denotes the European continental domain used for the CLM3.5 and eCLM simulations at 12.5 km resolution. The black boxes indicate the two regional domains for eCLM in biogeochemistry mode, covering the Iberian Peninsula (IP) and a region in Eastern Europe (EE), simulated at approximately 3 km resolution.}\label{fig:regions}
\end{figure}

We have 2 setups for the eCLM model: (1) The continental eCLM setup is applied over Europe (Fig.~\ref{fig:regions}, red boundary) on the EUR-11 grid at 12.5 km resolution in satellite phenology (SP) mode which uses prescribed vegetation phenology without explicitly simulating carbon-related variables. An overview of the continental-scale eCLM setup is provided in \cite{ewerdwalbesloh_grace_2026}. (2) The regional setup operates at a higher spatial resolution of approximately 3 km and covers two subregions of the European domain: the Iberian Peninsula and a region in Eastern Europe (Fig.~\ref{fig:regions}, black boundary). It uses the same input data sources as the continental setup, reprocessed for the higher-resolution grid and runs in biogeochemistry (BGC) mode, in which carbon and nitrogen cycles are simulated dynamically providing prognostic outputs of LAI and GPP. The carbon costs caused by the acquisition of plant nitrogen are accounted for by the Fixation and Uptake of Nitrogen model \citep{shi2016} and the Leaf Utilization of Nitrogen for Assimilation scheme \citep{ali2016} partitions leaf nitrogen to optimize photosynthesis capacity.

\subsection{Input simulated truth and error information}
\label{sec:truth}
To run the synthetic MAGIC assimilation experiments as OSSE (Sec.~\ref{sec:simulator}), we need to derive a simulated truth from a different model simulation that serves as observational input. Thus, our DA simulation experiments contain now two different model versions: 
\begin{enumerate}
    \item A model version used to represent the simulated truth. Synthetic observations are generated from this truth and are used as input into the assimilation.
    \item A model version underlying the data assimilation. Generally, the assimilation systems run a forward prediction and an update step for each time interval. A model simulation performs the forward prediction. The settings chosen for the model forward prediction are different than the settings chosen for the model version for the simulated truth.
\end{enumerate}
Both model versions differ systematically from each other and we summarize the settings for each model used in the four DA systems in Tab.~\ref{tab:settings}. For some DA experiments, we compare assimilation outputs towards the truth for evaluation.

For the global WaterGAP DA, the simulated truth is derived from a WaterGAP 2.2e simulation considering anthropogenic water use based on the W5E5 forcing data set \citep{lange2022}. The model underlying the WaterGAP assimilation is again a WaterGAP2.2.e version but neglects anthropogenic water use and is based on the ERA5 forcing data.  For both the truth and DA ensembles, we perform precipitation and temperature as well as the parameter perturbation following \cite{gerdener2024}. For the continental CLM3.5 DA experiment, the simulated truth is generated with eCLM forced with ERA5, whereas the assimilation system uses the CLM3.5 model driven by the regional COSMO-REA6 forcing. Conversely, for the continental eCLM DA experiment, the simulated truth is generated with the CLM3.5 model and COSMO-REA6 forcing, whereas the assimilation system uses eCLM forced with ERA5. Thus, the two experiments exchange the roles of model used for the simulated truth and assimilation model. For both the truth and DA ensembles, hydraulic properties and atmospheric forcings are perturbed to generate ensemble members, as described in \cite{ewerdwalbesloh_grace_2026}.

For the regional eCLM DA, both the truth simulation and the DA runs are conducted using eCLM with BGC mode enabled, driven by COSMO-REA6 forcing. To derive the synthetic truth, the eCLM is forced with data from 2002 to 2004. These synthetic observations are assimilated into DA runs forced with data from a different time period, i.e. relatively wetter conditions from 1996 to 1998, introducing a forcing mismatch between truth and assimilation, in contrast to the structural (model-based) differences also included in the global and continental experiments above. This setup focusing on two different time periods was chosen for the regional eCLM DA because including a new forcing data and model version would require an enormous amount on computational costs. Perturbations of atmospheric forcings and soil hydraulic properties for ensemble generation follow \cite{ewerdwalbesloh_grace_2026}.

As introduced before, compared with CLM3.5, eCLM includes a deeper soil column, improved hydraulic parameterizations, and revised runoff generation schemes. These model differences are expected to influence how TWS increments from the comparison of observation and model simulations developed during DA are distributed within the soil column and how they affect runoff generation. Both model configurations have previously been applied in real-world GRACE DA experiments and have shown good agreement with GRACE-derived TWSA and independent observation-based datasets in terms of variability and trends \citep[e.g.,][]{springer_evidence_2019, ewerdwalbesloh_grace_2026}.

Next, we generate monthly or 5-daily observations from the simulated truth by perturbing it with random errors of the mission scenarios. These errors are derived for GRACE-C-like and MAGIC mission configurations from the VCMs from the closed-loop simulator (Sec.~\ref{sec:simulator}). For the multivariate experiments in eCLM, daily surface (SWC) observations are generated from the simulated truth and perturbed with random errors assuming a standard deviation of 0.04 $mm^3/mm^3$. This value corresponds
to the target retrieval uncertainty of the SMAP mission \citep{entekhabi_soil_2010} and is consistent with observation error specifications adopted in previous SWC assimilation studies \citep[e.g.,][]{de_lannoy_assimilation_2016}. To reduce the computational cost of the assimilation experiments, 15.000 SWC observations are randomly sampled from the available observations at each daily assimilation time step.

\subsection{Data assimilation methods and settings}
\label{sec:DA_settings}
We compare TWSA DA outputs from four systems, which are run sharing the same filter algorithm but vary in the underlying model, its spatio-temporal resolutions, domain (global, continental, regional), assimilation frequency and observation treatment.

The global WaterGAP DA \citep{gerdener2023global,gerdener2024} water storage outputs on approximately 50 km are also published as Global Land Water Storage \citep[GLWS][]{gerdener2026gagp}. WaterGAP DA assimilates currently only monthly TWSA; the remaining assimilation systems additionally integrate the 5-daily TWSA product. The continental CLM3.5 system assimilates TWSA only \citep{springer_water_2019}, whereas the continental 12.5 km eCLM system over Europe (EURO-CORDEX domain) enables additionally the univariate assimilation of TWSA the assimilation of surface SWC in a multivariate setup. The regional eCLM DA on approximately 3 km assimilates TWSA only focusing the Iberian Peninsula and Eastern Europe.

In the multivariate continental eCLM system, TWSA and surface SWC can be assimilated sequentially. The individual analysis updates follow the procedures as described for TWSA assimilation in \cite{ewerdwalbesloh_grace_2026} and for SWC assimilation as described in \cite{sivaprasad_assimilating_2026}. SWC observations are assimilated daily at 13:00, whereas TWSA observations are assimilated at the end of each 5-day or monthly assimilation window, depending on the experiment. On days when TWSA is assimilated, the SWC update is performed first, followed by the TWSA update using the updated ensemble state. On the remaining days, only the SWC update is performed.

All systems use the Localized Error Subspace Transform Kalman Filter (LESTKF) \citep{evensen2003ensemble, nerger2012unification}, implemented via the Parallel Data Assimilation Framework (PDAF) \citep{nerger2013software,nerger2019efficient}. The LESTKF is a deterministic square-root ensemble filter that computes the analysis within the error subspace spanned by the ensemble perturbations and updates the ensemble through a symmetric square-root transformation. This avoids perturbing the observations while ensuring that the updated ensemble represents the Kalman analysis mean and covariance within the ensemble subspace \citep{nerger2012unification}. Localization is applied in observation space to restrict the influence of observations to the corresponding local analysis domains and to suppress spurious long-range correlations arising from the finite ensemble size. A common cut-off radius of 700 km for the observations within each local analysis domain is implemented throughout each of the assimilation runs. Inflation is applied in all continental assimilation systems to counteract the reduction of ensemble spread. A locally adaptive forgetting factor is computed for each analysis domain based on the ensemble spread in observation space and observation uncertainty. For TWSA, forgetting factors range between 0.95 and 1.00. In the multivariate assimilation runs, no inflation is applied to SWC.

As the LESTKF is an assimilation filter that requires an ensemble to represent model uncertainty, ensemble members are generated by perturbing the atmospheric forcing and model parameters, using the forcing data sets described in Section~\ref{sec:truth}. For the global WaterGAP DA, precipitation, temperature and model parameters are perturbed following \citet{gerdener2024}, resulting in 32 ensemble members.
The eCLM based assimilation setups use the soil hydraulic parameter and atmospheric forcing perturbation as described in \citet{ewerdwalbesloh_grace_2026}, with 32 ensemble members for the regional system and 50 for the continental eCLM DA. The continental CLM3.5 DA utilizes 64 ensembles members and perturbations according to \citet{springer_water_2019}. 

The input grid size of the observation and its error treatment differs between the global and continental/regional configurations. In the published GRACE/-FO WaterGAP DA data set (GLWS) the full observation error covariances of the spherical harmonic coefficients are propagated to a spatial 4$^\circ$ TWSA grid, preserving correlated observation error structures. Correlated observation errors in the real-world GRACE/-FO WaterGAP DA system are accounted for implicitly through domain localization \citep{gerdener2024}, where each local update is restricted using a 700 km cut-off radius for observations, which also reduces model correlations. Most GRACE/-FO data assimilation studies neglect observation correlations by only utilizing the diagonal error structure or uniform error information \citep{zaitchik_assimilation_2008,li_assimilation_2012,kumar2016assimilation}. The continental and regional DA setups assimilate observations of 2$^\circ$ spatial resolution, but rely on a diagonal error structure. In the multivariate setup, SWC observation uncertainties are prescribed using a uniform error of 0.04 $mm^3/mm^3$. The chosen assimilation settings are summarized in Table ~\ref{tab:settings}. 

\begin{table*}
\centering
\small
\caption{Settings of the synthetic truth models, the models underlying the assimilations and tuning of assimilations for the global, continental and regional TWSA data assimilation systems. All systems use the LESTKF with observation localization and a cut-off radius of 700 km. Entries marked with * apply to the multivariate continental eCLM DA only.}

\begin{tabular}{l | c | c c | c}
\hline
 & \textbf{Global} & \multicolumn{2}{c|}{\textbf{Continental}} & \textbf{Regional} \\
 & WaterGAP DA & CLM3.5 DA & eCLM DA & eCLM DA \\

\hline
\multicolumn{5}{l}{\textit{Synthetic truth}} \\
Model & \makecell{WaterGAP 2.2e\\(with water use)} & eCLM (SP) & CLM3.5 & eCLM (BGC) \\
Forcing & W5E5 & ERA5 & COSMO-REA6 & \makecell{COSMO-REA6\\(2002--2004)} \\

\hline
\multicolumn{5}{l}{\textit{Assimilation model}} \\
Model & \makecell{WaterGAP 2.2e\\(no water use)} & CLM3.5 & eCLM (SP) & eCLM (BGC) \\
Domain & Global over land & EURO-CORDEX & EURO-CORDEX & \makecell{Iberian Peninsula\\ Eastern Europe} \\
Grid & 0.5$^\circ$ & 12.5 km & 12.5 km & 3 km \\
Forcing & ERA5 & COSMO-REA6 & ERA5 & \makecell{COSMO-REA6\\(1996--1998)} \\

\hline
\multicolumn{5}{l}{\textit{Assimilation settings}} \\
Observations & TWSA & TWSA & TWSA (+ SWC*) & TWSA \\
TWSA interval & monthly & \makecell{monthly or\\5-daily} & \makecell{monthly or\\5-daily} & \makecell{monthly or\\5-daily} \\
SWC interval & -- & -- & daily* & -- \\
Observation grid & 4$^\circ$ & 2$^\circ$ & 2$^\circ$ & 2$^\circ$ \\
Observation errors & full covariance & diagonal & diagonal & diagonal \\
Ensemble size & 32 & 64 & 50 & 32 \\
Inflation & No & adaptive & adaptive & No \\
\hline
\end{tabular}
\label{tab:settings}
\end{table*}


\section{Results}
We want to provide a broad perspective of global, continental and regional GRACE/-FO assimilation systems and the potential benefit that could be gained from the MAGIC twin-pair constellation for these. Therefore, we will introduce a short review and limitations for existing global, continental, and regional DA systems. We will introduce exemplary applications for each of the four DA systems to demonstrate the benefit of MAGIC. For the global WaterGAP DA system, we will investigate how errors of trends, seasonal amplitudes and drought monitoring could improve when extending the existing GRACE/-FO DA dataset \citep[GLWS][]{gerdener2026gagp} by a MAGIC DA record, and thus deriving a longer time series of water storages from assimilation.

For the continental CLM3.5 and eCLM DA systems, we will evaluate the inter-annual performance of the framework on catchment scale for the European CORDEX domain, assess the impact of monthly and 5-daily DA on simulated fluxes in the case of one exemplary extreme event. Using both DA versions allows us to investigate how the impact of data assimilation depends on model complexity and process representation. For both DAs, we conduct experiments using consistently simulated GRACE-C-like and MAGIC observation error covariance matrices at monthly and 5-daily temporal resolutions, with GRACE-C-like assumed to approximately represent the observation characteristics of the currently operational GRACE-FO mission. With this setup, we can isolate the effects of the enhanced spatial and temporal resolution of the MAGIC double-pair constellation compared with GRACE-C-like within a controlled twin experiment. In addition, we conduct multivariate eCLM DA experiments in which MAGIC TWSA and surface soil moisture are assimilated jointly, providing insights into the effects of multivariate assimilation approaches for future gravity missions.

For the regional eCLM DA system, we will assess how increasing the assimilation frequency from monthly to 5-daily, as enabled by MAGIC, impacts not only the simulated TWSA on the fine 3 km resolution but also the representation of GPP and LAI. Since TWSA DA influence vegetation only indirectly through updates in root-zone water availability, more frequent updates could correct the soil water estimates closer to the time scales on which it affects photosynthesis.

\subsection{Global-scale assimilation}
\subsubsection{Review and limitations of existing global GRACE/-FO assimilation frameworks}
Since the first GRACE data assimilation framework in 2008 \citep{zaitchik2008assimilation} more than a dozen GRACE and GRACE-FO data assimilation systems at regional or continental scales have been developed. These studies elaborated the great potential of assimilating TWSA from gravity satellite observations into land surface models or hydrological models for improving applications of the hydrological water cycle. A decade ago no global-scale assimilation system existed but, meanwhile, five global studies have investigated the assimilation of GRACE and GRACE-FO TWSA into hydrological and land surface models. \cite{li2019global} and \cite{felsberg_global_2021} assimilated TWSA derived from GRACE/-FO Mass Concentration (mascon) solutions into the Catchment Land Surface Model (CLSM), a global land surface modeling framework developed by NASA. Similarly, \cite{khaki2023assimilation} incorporated globally gridded GRACE/-FO TWSA products derived from mascon solutions or K-band inter-satellite ranges, into the World-Wide Water Resources Assessment (W3RA) model. \cite{forootan2024global} introduced a novel Markov Chain Monte Carlo Data Assimilation (MCMC-DA) framework to assimilate globally gridded GRACE/-FO TWSA into the W3RA model. Our own global-scale assimilation system assimilated TWSA from GRACE/-FO into the WaterGAP global hydrological model \citep{gerdener2023global,gerdener2024}. While most global and regional assimilation studies neglect observation correlations or apply a spatially uniform  error, we include them as derived from the satellite geopotential retrieval.

In fact, all mentioned global assimilation systems come with different advantages and difficulties, for example, while the CLSM assimilation is able to derive soil layers in different depths, the WaterGAP assimilation has only one root-zone soil layer but has an improved representation of groundwater. However, a few challenges apply to all global assimilation systems. The existing global (and continental and regional) real-world assimilation systems have a limited time span when the end of the GRACE-FO mission is reached. With the launch of MAGIC, the assimilation time series could be extended in the future, ensuring the continuation of not only TWSA but also other storages and fluxes such as soil moisture, groundwater, surface water, discharge, evapotranspiration from assimilation. The spatial resolution of real GRACE/-FO TWSA is currently limited to about 300 km, with an improvement in spatial resolution one could not only enhance the TWSA spatial resolution prior to assimilation but also improve realism of spatially downscaled storages and fluxes after assimilation. Thus, global MAGIC assimilation systems might come with higher precision of water storages as compared to the current real world GRACE/-FO data assimilation systems improving uncertainty estimates of variables of the hydrological water cycle. 

Thus, we will extend the real GRACE/-FO global WaterGAP DA from 2003 to 2019 \citep{gerdener2024} with a synthetic MAGIC WaterGAP DA that will be provided here for 2020 to 2035. The synthetic MAGIC DA (setup described in Sec.~\ref{sec:simulator}) will run with similar DA settings as the GRACE/-FO DA and with hindcast forcing data. After projecting this retrospective synthetic global MAGIC DA to the time following the GRACE/-FO DA, we can derive continued DA water storages. We will investigate in the following experiments in how far the extension of the time period and the expected improved uncertainty of MAGIC assimilation affects the estimation of storage errors for linear trend, annual and semi-annual amplitude, and drought monitoring. This means we can investigate errors in signatures such as linear trends or annual amplitudes either for 17 years of GRACE/-FO DA era only and compare them to signature errors that can be derived from 33 years of data derived from the extending GRACE/-FO DA with MAGIC DA on the global scale.

\subsubsection{Impact of TWSA extension through MAGIC assimilation on trend and amplitude errors}
\begin{figure}[h!]
\begin{center}
\includegraphics[width=18cm,trim= 0.5cm 0.5cm 0.5cm 0.5cm, clip]{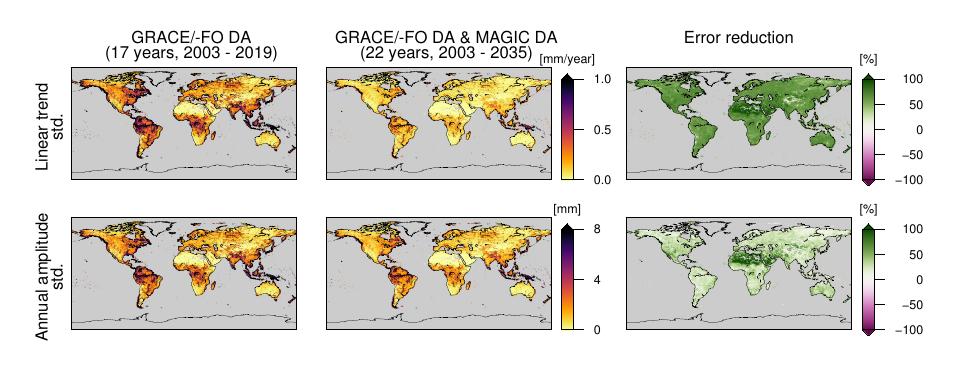}
\end{center}
\caption{Standard deviations of linear trends (top) given in mm/year or annual amplitudes (bottom) given in mm either derived for 2003 to 2019 from real global GRACE/-FO DA only (left) or for 2003 to 2035 from the extension of the real GRACE/-FO DA with the synthetic MAGIC DA (middle) and the relative error improvement [\%] of standard deviations of long time series (GRACE/-FO DA extended with MAGIC DA) towards the short time series (GRACE/-FO DA) right). Green indicates smaller trend or amplitude errors of GRACE/-FO extended with MAGIC DA than for the GRACE/-FO DA only.}\label{fig1:trendAnnErrors}
\end{figure}

We first focus our analysis here on the errors of TWSA trends as derived from the real-world global GRACE/-FO WaterGAP assimilation (Fig.~\ref{fig1:trendAnnErrors}, top left), i.e. a total of 17 years before analyzing the impact that a temporal extension with MAGIC DA would have. To derive the errors, we use standard deviations estimated from the DA ensemble spread for each TWSA month and apply variance propagation through the trend estimation. As expected, trend errors derived from GRACE/-FO DA are highest in humid regions, for example, in Brazil or India with the Amazon and Ganges river basins, while being lower in arid regions such as in South Africa or Australia. Next, we extend TWSA of the real GRACE/-FO DA with the synthetic MAGIC DA and analyze errors in TWSA trends as derived from the long time series of 33 years (Fig.~\ref{fig1:trendAnnErrors}, top middle). The global mean (spatial average of all land 0.5° grids) trend error of 17 years GRACE/-FO assimilation outputs clearly reduces from 0.57 mm per year to 0.2 mm per year. More specifically, the basin averaged (of 0.5° grids) trend error is strongly reduced by 0.90, 0.53, 0.72 mm per year for the Amazon, Ganges, and Congo river basins, which represents a relative error reduction of about 63, 67, and 67\%, respectively. Fewer absolute reduction is found for the Murray-Darling and Danube river basins with 0.23 mm per year improvement for both, which we attribute to the trend error in GRACE/-FO DA being already much lower as compared to the trend errors in the Amazon, Ganges, and Congo river basins. However, the relative error improvement is in the same magnitude with 64 and 63\%. On global average, we derive an improvement of 65\% in linear trend errors by having a longer time series available and by having lower errors from MAGIC DA.

When repeating the analysis on an 1.5° or an 3° grid, we find a similar percentage improvement for linear trend errors derived from the long time series as compared to the short time series of about 65\%. However, the total magnitude of trend errors is -- as expected -- lower as compared to the 0.5° grids. For example, on global average of 3° grids, using 17 years of GRACE/-FO DA provides a trend error of 0.38 mm/yr and the 33 years of GRACE/-FO DA prolonged with the MAGIC DA provides a TWSA trend error of 0.13 mm/yr.

When analyzing errors of annual TWSA amplitudes (Fig.~\ref{fig1:trendAnnErrors}, bottom row), we also find reductions in errors when extending the GRACE/-FO DA 17-year era to the combined GRACE/-FO and MAGIC DA 33 years. The global mean annual amplitude error of 0.5° grids in GRACE/-FO DA is reduced from 3.83 mm to 2.60 mm when extending with MAGIC DA, which is an improvement of about 32\%. Thus, we find a major improvement in annual amplitude errors when extending the assimilation time series, although the percentage improvement is not as high as for the trend errors. A strong reduction in annual amplitude error is found, for example, for the Amazon basin, Ganges, and Congo basins with reductions by 27, 30, and 32\%, respectively. When repeating the analysis for semi-annual amplitude errors (results not shown), we can find very similar improvement values as for annual amplitude errors. 

In summary, by extending the real GRACE/-FO DA time series with synthetic MAGIC DA, we derive an improvement in linear trend, annual amplitude, and semi-annual amplitude errors for TWSA with highest percentage of improvements found for the trend errors on the global average.

\subsubsection{Impact of groundwater, soil moisture and surface water extension through MAGIC assimilation on trend and amplitude errors}
DA enables us to not only derive improvements in TWSA, but also vertically disaggregate the TWSA into water storage compartments from WaterGAP DA, i.e. groundwater, surface water (lakes, wetlands, rivers, reservoirs), root-zone soil moisture, etc. Thus, we can also analyze whether the found error improvements in TWSA from the extending the real GRACE/-FO DA with synthetic MAGIC DA also transplant to other compartments. We compare now errors of different compartments at the example of linear trends. As described above, the global improvement of linear trend errors is about 65\% for TWSA. Although the absolute errors of groundwater, surface water and soil moisture from the 17-years GRACE/-FO DA are generally lower than the TWSA trend error (0.28, 0.42, and 0.16 mm per year), the percentage improvement is in a similar magnitude as for TWSA with 68, 64, and 62\% when extending the 17 years GRACE/-FO DA with the MAGIC DA to 33 years. Interestingly, the global mean (0.5° grids) absolute trend error of surface water storage is larger than the error of groundwater and soil moisture. We attribute this to the fact that river, lakes, wetlands and reservoirs are updated independently during assimilation and afterwards these layers are aggregated to derive surface water, which could generate a bigger uncertainty than having the uncertainty of one single layer. For groundwater and soil moisture no aggregation is performed and the errors are representative for a single layer. 

\begin{figure}[h!]
\begin{center}
\includegraphics[width=13cm,trim= 0cm 12.8cm 4cm 0cm, clip]{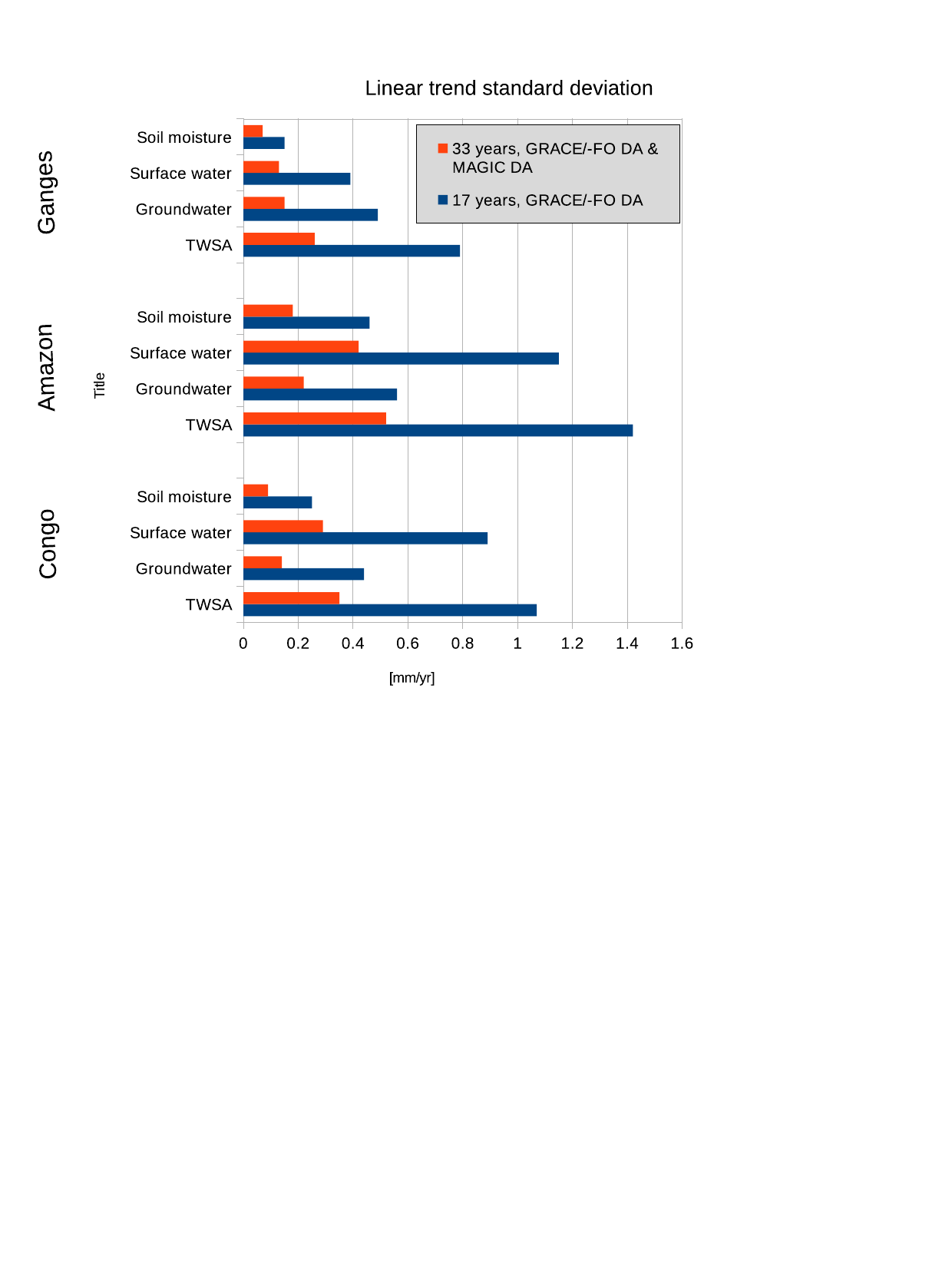}
\end{center}
\caption{Storage standard deviations [mm/year] of linear trends for the Congo, Amazon, and Ganges river basins either derived from 17 years (2003 to 2019) of real GRACE/-FO DA only (blue) or from 33 years (2003 to 2035) from the temporal extension of GRACE/-FO DA with the synthetic MAGIC DA (orange). The standard deviations are derived as basin average of 0.5° grid cells.}\label{fig2:BasinTrendErrors}
\end{figure}

Next, we exemplarily show trend errors for a few large river basins (the Ganges, Amazon, and Congo basin, Fig.~\ref{fig2:BasinTrendErrors}) situated in different climate and vegetation regimes. We derive improvements in basin-averaged (from 0.5° grids) trend errors for all three basins for groundwater, surface water, and soil moisture when extending the 17 years GRACE/-FO DA to the 33 years with the MAGIC DA. Similar as for the global analysis, surface water trend errors are generally higher than groundwater and soil moisture trend errors for the Amazon and Congo basin. For example, with 33 year of GRACE/-FO DA the trend error in the Amazon basin is about 1.2 mm per year for surface water while it is about 0.6 mm per year for groundwater. In contrast, for the Ganges basin, the groundwater trend errors are higher than the surface water and soil moisture trend errors. We hypothesize that the groundwater trend errors could be strongly affected by anthropogenic water use in the Ganges basin because groundwater use is very common there \citep[e.g.,][]{LAVETI2021125800}. Similar differences between the Amazon and Congo basin towards the Ganges basin can also be drawn when analyzing annual and semi-annual amplitudes for the storages. As it also was the case for TWSA, the global mean improvement is generally lower for annual and semi-annual amplitudes than for trends but still significant with around 30\%. For example, the annual amplitude error for groundwater improves by 34\% when using the GRACE/-FO DA extended with MAGIC DA instead of GRACE/-FO DA only, while the improvement reduces slightly to 25\% for soil moisture annual amplitude errors. 

All in all, we found strong improvements in errors for trend estimation and seasonal climatology across storages on a global and basin scale due to the temporal extension of the GRACE/-FO DA with MAGIC DA leading to 33 years of data and due to the higher precision in MAGIC DA.

\subsubsection{Changes in drought severity from total water storage extension with MAGIC assimilation}
\begin{figure}[h!]
\begin{center}
\includegraphics[width=18cm,trim= 0cm 1.5cm 0cm 0cm, clip]{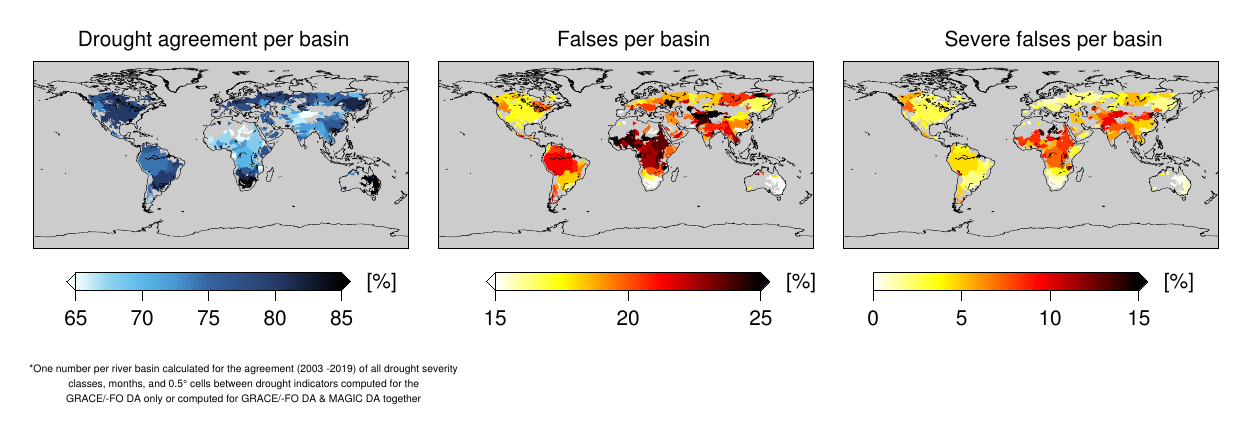}
\end{center}
\caption{Drought agreement (left), drought falses (middle), and severe drought falses (right) per basin for the overlap period 2003 to 2019 from computing DSI either based on 17 years of global GRACE/-FO DA only or based on 33 years global GRACE/-FO DA extended MAGIC DA. The agreement is computed as a percentage of all spatial 0.5° grid and months from 2003 to 2019 that indicate the same severity classes as estimated with DSI. If severity classes differ by one class we define that as falses and by two or more classes as severe falses.}\label{fig3:DroughtAgreement}
\end{figure}

For monitoring drought with GRACE/-FO DA, drought indicators are typically used \citep{mukhawana2023, gerdener_framework_2020}. These drought indicators assign single months of a time series to specific drought severity classes ranging from moderate to exceptionally dry. We will use the commonly applied Drought Severity Index \citep[DSI][]{zhao_global_2017} to monitor drought severity either when applying DSI computations to TWSA from 2003 to 2019 for GRACE/-FO DA only (17 years) or for the extended time period from 2003 to 2035 for GRACE/-FO DA and MAGIC DA jointly (33 years). Thus, the extended GRACE/-FO DA with MAGIC DA meets the often suggested length of data of at least 30 years \citep{svoboda2016handbook} for drought computation. Next, we compare both DSI computations for the overlapping period 2003 to 2019 and extract agreement or disagreements in drought severity (Fig.~\ref{fig3:DroughtAgreement}) shown for the 140 largest river basins. Agreement is defined as percentage of all 0.5° grids and months from 2003 to 2019 in the respective basin that agree in the exact same drought severity class. The average agreement for all 140 basins is about 75\%. We find that no basin has an agreement in drought severity of 90\% or higher (Fig.~\ref{fig3:DroughtAgreement}, left). Thus, using 33 instead of 17 years provides a much better drought statistic, from which we get a more robust drought severity identification worldwide. Nonetheless, we still find many basins with a high agreement between 85 and 90\%, for example, for the Burdekin, Orange-Senqu, and Alabama river basins. Basins that experience an agreement of 40\% or lower are not found. We identify several basins with a medium agreement, for example, the Nile and Indus basins with agreements of about 68 and 66\%, respectively.

We can also analyze how much the severity classes during 2003 to 2019 disagree computed from a short time period of 17 years from real GRACE/-FO DA compared to when computed from a long time period of 33 years from GRACE/-FO DA extended with the synthetic MAGIC DA. The disagreement is defined for two cases: 1) the two DSI time series disagree by one drought severity class (falses) or 2) they disagree by two or more severity classes (severe falses) during the overlap. The severe falses (Fig.~\ref{fig3:DroughtAgreement}, right) are concentrated in West/Central Africa and South Asia, for example, the Kufra, Rovuma, Tarim and Hai He river basins although these severe falses occur only in up to 20\% of all grids and months per basin. Averaging the percentage of severe falses across the 140 basins yields a mean value of about 4.5\%. For falses the mean across the basins is about 20\%. Thus, severely falses are -- as expected -- much more rare than falses. Basins experiencing one of the largest percentage of falses are, for example, the Rovuma, Tarim, and Ogooue basins with about 27\%.

Concluding, we find that the agreement in drought severity from global assimilation between a DSI computed from the GRACE/-FO DA era and from the GRACE/-FO DA extended MAGIC DA era is for most basins worldwide high during the overlap time period 2003 to 2019 but strongly depends on the respective basin. In many basins, we can improve drought statistics with the GRACE/-FO DA extended with MAGIC DA resulting in 33 years of baseline period for index computation. The disagreement is most apparent in Central and West African and South Asian river basins and, thus, the long time period would improve drought statistic especially here.

\subsection{Continental-scale assimilation over Europe}

\subsubsection{Review and limitations of existing continental GRACE/-FO assimilation frameworks}
Continental-scale assimilation of GRACE TWSA observations has received considerably less attention than studies focusing on individual river basins. Early continental-scale applications include experiments over North America using CLM3.5 at 4° resolution and CLSM for snow and drought monitoring, respectively \citep{su_multisensor_2010,houborg_drought_2012}. \citet{kumar_assimilation_2016} later assimilated gridded GRACE observations within the North American Land Data Assimilation System (NLDAS) at 1/8° resolution. More recently, multivariate assimilation of GRACE-FO TWSA and SMAP soil moisture into Noah-MP at 1/8° resolution has been used to generate high-resolution daily groundwater estimates across the United States \citep{ghaneei_effective_2026}.

For Europe, \citet{li_assimilation_2012} conducted one of the first continental-scale GRACE assimilation experiments using CLSM to evaluate its potential for drought monitoring in western and central Europe. Subsequently, \citet{springer_evidence_2019} and \citet{klos_quantifying_2021} applied GRACE DA with CLM3.5 over Europe at 0.11° resolution, primarily to improve estimates of hydrological loading relevant for GNSS observations. More recently, \citet{ewerdwalbesloh_grace_2026} investigated GRACE TWSA assimilation in eCLM over Europe, focusing on the role of state vector design and model uncertainty representation in controlling assimilation performance.

Additional continental-scale studies have been performed over Australia and South America. \citet{tian_improved_2017} assimilated GRACE terrestrial water storage together with SMOS soil moisture into the W3RA hydrological model over Australia at 0.5° resolution. \citet{khaki_assessing_2017} evaluated different sequential assimilation techniques for integrating GRACE data into W3RA, while \citet{khaki_accounting_2017} investigated the role of spatially correlated GRACE errors through localization within the same framework. \citet{khaki_nonparametric_2018} subsequently introduced a nonparametric Kalman-Takens filter for continental-scale hydrological assimilation over Australia. \citet{tangdamrongsub_evaluation_2018} evaluated groundwater storage variations from GRACE assimilation with PCR-GLOBWB for Australia. Finally, \citet{khaki_assessing_2019} extended the Australian W3RA framework to South America using multi-mission satellite assimilation at 5-day temporal resolution.

Together, these studies demonstrate that continental-scale GRACE data assimilation can improve estimates of snow, drought conditions, groundwater storage, and hydrological loading. However, the existing frameworks differ substantially in model structure, assimilation algorithms, GRACE products, and the treatment of observation and model uncertainty, making it difficult to isolate the impact of individual design choices. In particular, previous European studies have focused either on drought monitoring \citep{li_assimilation_2012}, hydrological loading applications \citep{springer_evidence_2019,klos_quantifying_2021}, or methodological aspects of the eCLM assimilation framework \citep{ewerdwalbesloh_grace_2026}. A systematic comparison of different gravity mission covariance structures, monthly versus submonthly assimilation frequencies, and the influence of different model versions has not yet been carried out for Europe. 

Thus, MAGIC CLM3.5 DA and eCLM DA could potentially not only enhance monthly but also submonthly continental water storage and fluxes and uncover benefits compared to a GRACE-C-like DA (representative for GRACE-FO). 

The following experiments in this section show the performance of MAGIC DA for (i) the CLM3.5 DA framework across Europe, (ii) the April 2006 Danube flood using both CLM3.5 and eCLM, and (iii) the vertical downward propagation of the 2007 Danube drought through the soil column under univariate (TWSA-only) and multivariate (TWSA and surface soil moisture) DA with eCLM.

\subsubsection{Impact of MAGIC assimilation on TWSA estimates}
\begin{figure}[h!]
\begin{center}
\includegraphics[width=15cm,trim= 0cm 0cm 0cm 0cm, clip]{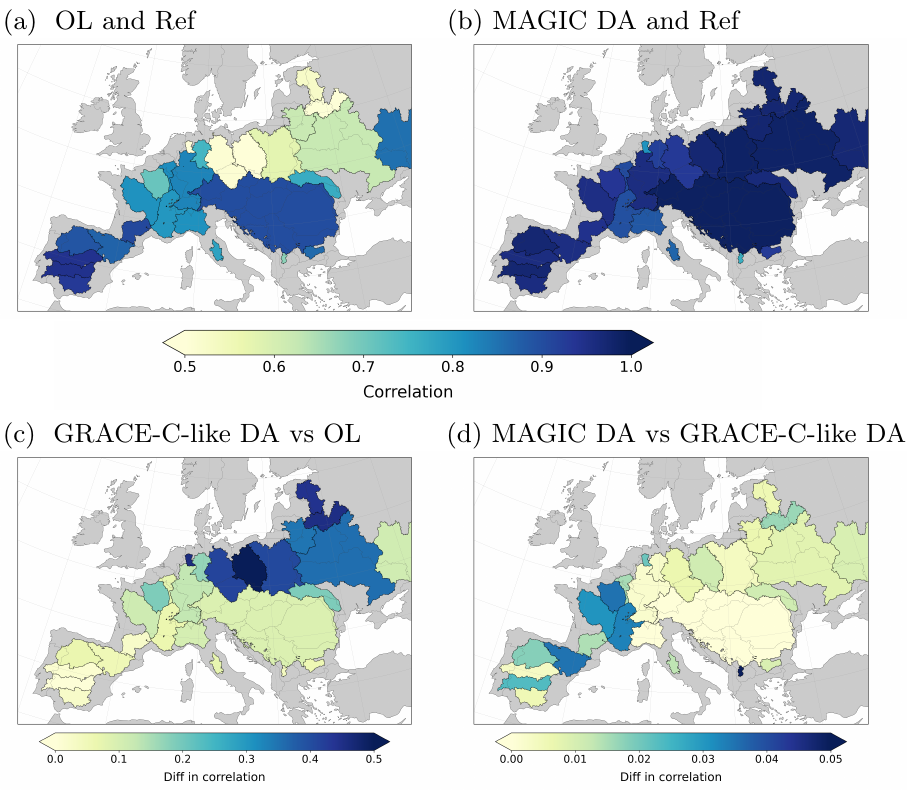}
\end{center}
\caption{First row shows the correlation between CLM3.5 OL and continental MAGIC CLM3.5 DA with respect to the reference truth from eCLM for monthly terrestrial water storage anomaly (TWSA) time series where trend and annual signal were removed (residual TWSA time series). The second row shows differences in correlation, with blue indicating improvements, for (c) change in correlation when using GRACE-C-like DA instead of OL and (d) additional correlation improvement from assimilating the MAGIC instead of GRACE-C-like scenario.}\label{fig4:clm3_correlations}
\end{figure}

Overall, the continental MAGIC CLM3.5 DA shows improved agreement with the eCLM-based reference truth of TWSA spatially averaged (from 12 km grids) across European catchments in terms of trends, Root Mean Square Difference (RMSD), and correlation compared with the open loop (OL) simulations (see SI, Figures S1, S2, S3, and Fig.~\ref{fig4:clm3_correlations}). TWSA differences between the CLM3.5 OL simulations and the eCLM reference are expected due to differences in set up (Sec.~\ref{sec:truth}). Data assimilation reduces these differences, with trend estimates improve particularly over the Iberian Peninsula, where the OL simulation shows trends of opposite sign, and over Central Europe, where assimilation reduces the discrepancy between the overly strong positive trends in the OL simulation and the synthetic reference due to the mentioned simulation differences (Sec.~\ref{sec:truth}). TWSA RMSD with respect to the reference truth decreases by approximately 20–40\% for the catchment-averaged TWSA when assimilating GRACE-C-like compared with the OL simulation, with the largest reductions occurring in Eastern Europe. Using the MAGIC DA instead of the GRACE-C-like DA provides an additional reduction of the RMSD by 3–5\% (Fig.~S2).

For the GRACE-C-like DA experiment, correlations between the assimilated CLM3.5 TWSA and the reference truth are particularly improved over the OL case in northeastern Europe, where assimilation increases correlations by approximately 0.3–0.4 (Fig.~S1). This improvement reflects the better representation of TWS variability, including the correction of underestimated seasonal amplitudes in the open-loop simulation. MAGIC DA further increases correlations compared with GRACE-C-like DA, although the additional improvement is smaller (up to 0.02–0.03). Fig.~\ref{fig4:clm3_correlations} illustrates correlations of deseasonalized and detrended TWSA for the OL and assimilation experiments. The largest assimilation-induced improvements occur over Eastern Europe, while the additional benefit of MAGIC compared with GRACE-C-like is particularly apparent over Western Europe, where correlations further increase by 0.03–0.04.

All in all, continental MAGIC CLM3.5 DA improves TWSA representation for varying metrics as compared to GRACE-C-like DA scenarios.

\subsubsection{Evaluating assimilation frequency and multivariate assimilation with MAGIC: The 2006 Danube Flood}
In 2006, the Danube catchment experienced two major flood events \citep[e.g.,][]{mikhailova_extreme_2012}. The first and more severe event occurred in April and was driven by rapid snowmelt combined with precipitation, affecting large parts of the upper and middle Danube basin, including Germany, Austria, Slovakia, and Hungary. A second flood wave occurred in June following intense precipitation, with particularly high water levels in downstream regions such as Romania and Bulgaria. These two events provide contrasting conditions to assess how assimilation of TWSA affects the simulated partitioning between storage changes and runoff during rapidly evolving hydrological events.

Here, we evaluate (1) the continental MAGIC CLM3.5 DA experiments at different temporal resolutions to evaluate the potential of short-term solutions, and (2) comparing the MAGIC CLM3.5 DA with MAGIC eCLM DA experiments, with additional synthetic surface soil moisture observations to elaborate the use of multivariate experiments. All scenarios are evaluated for spatial averages (from 12 km grids) over the Danube catchment. Runoff simulated by land surface models can exhibit substantial biases compared with gauging data, and differences between models can arise from their different runoff generation schemes. We are therefore interested not in the absolute differences in runoff between the two models, but in the temporal evolution of basin-averaged runoff and how it is affected by TWSA DA or multivariate DA.

\begin{figure}[h!]
\begin{center}
\includegraphics[width=18cm,trim= 0cm 0cm 0cm 0cm, clip]{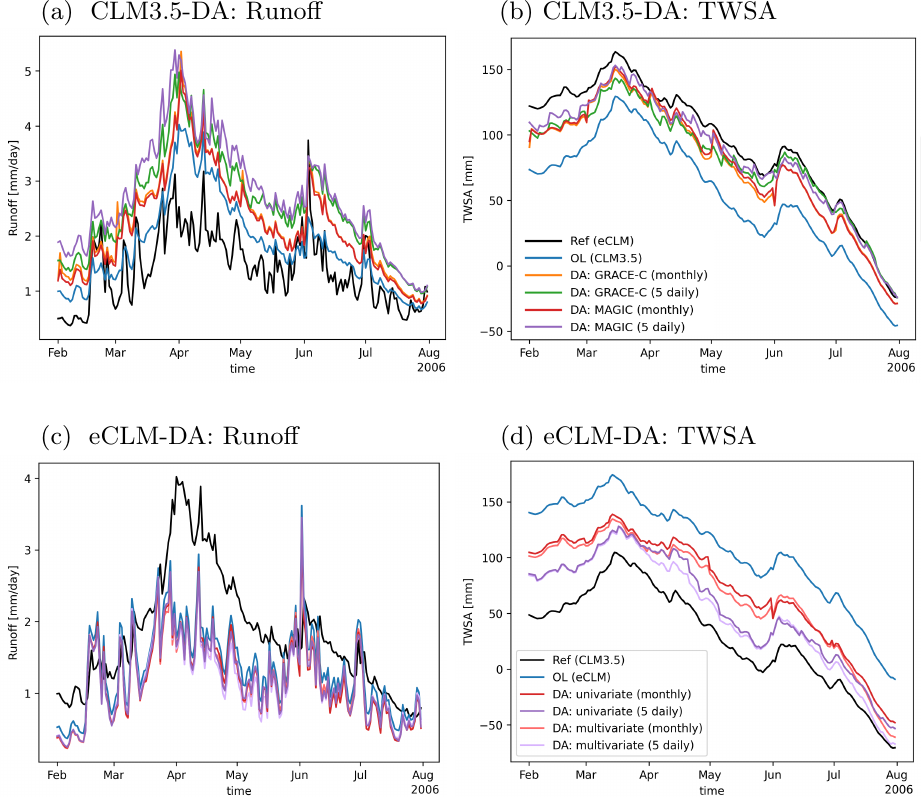}
\end{center}
\caption{(a) and (c) show surface and subsurface runoff, and (b) and (d) show TWSA in the Danube catchment in 2006 for the continental GRACE-C-like and MAGIC CLM3.5 DA (top row) and for the continental MAGIC eCLM DA in univariate or multivariate mode (bottom row).}\label{fig:clm_runoff}
\end{figure}

The eCLM OL exhibits larger TWSA than the CLM3.5 OL during both events maybe related to larger soil depth. In general, assimilation of the monthly and 5-daily TWSA increases TWSA in CLM3.5 DA, and these storage corrections subsequently propagate into the simulated runoff response, leading to increased discharge during both flood peaks  (Figure~\ref{fig:clm_runoff} (a) and (b)). This again demonstrates that TWSA-based assimilation not only adjusts TWSA but also affects the simulation of hydrological fluxes, which is consistent with other studies \citep[e.g., ][]{li_assimilation_2012}. The same general response is found for eCLM DA (Figure~\ref{fig:clm_runoff} (c) and (d)), although the sign and magnitude of the response differ. Here, assimilation reduces TWSA relative to the OL simulation, resulting in lower simulated discharge during both flood events.

For CLM3.5 DA, monthly assimilation produces a similar overall response; however, the 5-daily experiments lead to slightly larger increases in TWSA and substantially stronger runoff responses. In particular, the MAGIC 5-daily assimilation results in almost twice the runoff change compared with monthly assimilation. While monthly assimilation increases runoff by approximately 0.2–0.5 mm/day, the corresponding increase reaches 0.5-1.5 mm/day for the MAGIC 5-daily experiment. The stronger response is related to larger increments introduced by the MAGIC DA, particularly during the April flood event. In addition, the higher temporal resolution shifts the runoff response, with the 5-daily assimilation reaching the flood peak earlier than the corresponding monthly experiment. These results demonstrate that the temporal resolution of the assimilated MAGIC information influences not only the magnitude of storage corrections but also how rapidly these corrections propagate into hydrological fluxes during extreme events.

For the MAGIC eCLM DA in univariate or multivariate mode, the impact of TWSA DA on discharge is considerably smaller. The 5-daily multivariate MAGIC DA produces the largest deviation in discharge from the OL simulation, but the overall changes in discharge remain small compared with those obtained for the MAGIC CLM3.5 assimilation experiments. In particular, the timing of the flood peaks, including the June event, is not substantially affected by the assimilation. This weaker runoff response is consistent with the deeper soil column of eCLM, which provides a larger storage capacity and allows TWS corrections to be distributed over more and deeper layers. Consequently, the resulting changes in near-surface saturation and subsurface drainage are smaller, reducing the response of both runoff components.

The MAGIC eCLM assimilation in multivariate mode, which combines TWSA and surface soil moisture observational input, slightly improves TWSA agreement with the reference compared with the corresponding univariate eCLM experiments. This is notable because univariate assimilation often outperforms multivariate assimilation for an individual assimilated variable \citep{li_land_2024}. Here, the further improvement may result from the similar process representation in eCLM model version during assimilation and the CLM3.5 reference, such that constraining soil moisture towards the reference state also leads to a more similar TWSA state. This may be specific to the synthetic reference setup and is not necessarily expected for real-world experiments.

\subsubsection{Soil moisture propagation during drought conditions from MAGIC assimilation scenarios}

\begin{figure}[h!]
\begin{center}
\includegraphics[width=18cm,trim= 0cm 0cm 0cm 0cm, clip]{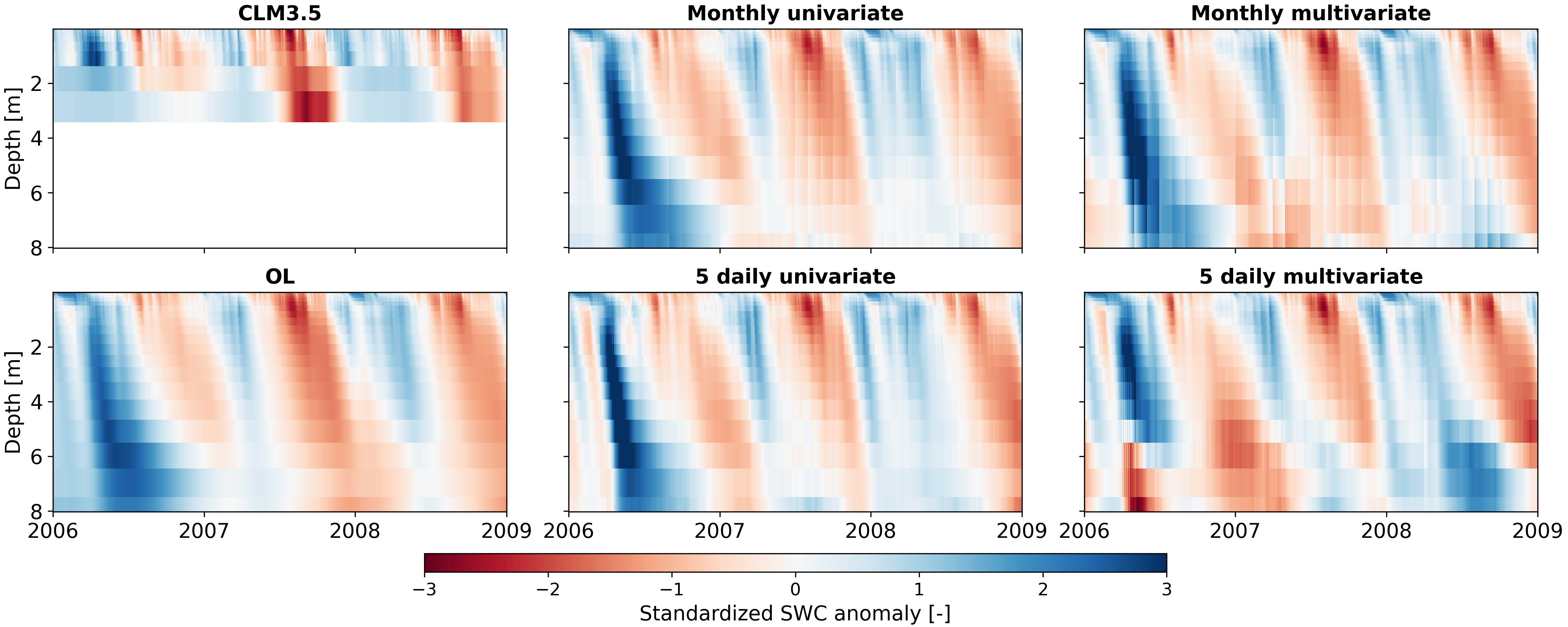}
\end{center}
\caption{SWC anomalies at different depth layers spatially averaged for the Danube catchment from 2006-2008. Anomalies are relative to 2003-2010 for each experiment. Shown are the reference truth CLM3.5 (top left), the eCLM OL (bottom left) and continental MAGIC eCLM DA experiments: univariate monthly (top center), univariate 5 daily (bottom center), multivariate monthly (top right), and multivariate 5 daily (bottom right). Negative values indicate drier-than-baseline conditions, while positive values indicate wetter-than-baseline conditions. Univariate assimilation refers to assimilating only TWSA observations whereas in multivariate assimilation, soil moisture is assimilated alongside TWSA.}\label{fig:propagation}
\end{figure}

To investigate the propagation of soil moisture anomalies through the soil column and the potential benefits of high-temporal-resolution MAGIC observations, Figure~\ref{fig:propagation} show standardized SWC anomalies spatially averaged (from 12 km grids) for the Danube catchments from 2006 to 2008. We focus here on the continental MAGIC eCLM assimilation scenarios, comparing univariate TWSA assimilation with the multivariate assimilation of TWSA and surface soil moisture. The anomalies are calculated separately for each soil layer using the mean and standard deviation over the full 2003–2010 simulation period. This normalization accounts for the substantially different variability between layers and allows the temporal evolution of soil moisture to be compared across the soil column despite the increasing layer thickness and generally lower variability at depth.

The OL simulation derived from eCLM shows a clear downward propagation of the dry conditions developing during 2007. Negative anomalies first emerge in the upper soil and subsequently extend into deeper layers, while the preceding wet conditions in 2006 are gradually replaced by drier conditions. The deeper layers retain the anomaly for longer and recover more slowly, illustrating the longer memory of soil moisture at depth. The much shallower soil column of the CLM3.5 truth simulation provides a clear contrast to eCLM OL. In CLM3.5, the dry anomaly reaches the bottom of the soil column considerably earlier and subsequently recovers faster. The deeper soil column of eCLM allows the anomaly to propagate over a larger depth range and persist for longer, particularly in the deepest layers. This difference illustrates how the vertical extent of the model soil column affects the timescale over which (assimilated) storage changes can influence soil moisture.

The propagation of soil water found in the OL simulation is also apparent in the assimilation experiments, although its structure depends on the assimilation frequency. The difference between monthly and 5-daily MAGIC eCLM assimilation is most apparent in the temporal evolution and depth-wise propagation of the anomalies. The 5-daily experiments exhibit a more gradual downward propagation of the dry anomaly over time, whereas the monthly experiments show a more fragmented vertical structure. The higher temporal resolution therefore allows the assimilation system to modify the soil moisture state more frequently during the development of the dry period, resulting in a clearer propagation of the anomaly through the soil column. Additionally, the 5-daily univariate experiment also responds somewhat earlier to the transition towards wetter conditions in 2006.

These results illustrate the potential benefit of future high-temporal-resolution TWSA observations. Compared with monthly MAGIC assimilation, more frequent 5-daily updates provide additional information during the development and recovery of rapidly evolving soil moisture anomalies and result in not only a different timing of soil moisture peaks but also a more continuous propagation of the assimilated signal through the soil column. However, the magnitude of these effects depends on the assimilation system and should not be directly interpreted as a prediction of the impact of a future MAGIC mission.

Differences between univariate and multivariate MAGIC eCLM assimilation are comparatively subtle for the vertical soil water propagation. The 5-daily multivariate experiment nevertheless shows some distinct vertical structures, particularly during the initial development of the dry period, indicating that the additional soil moisture constraint modifies the distribution and subsequent evolution of the assimilated storage changes in some periods but not fundamentally changing its overall propagation. These limited differences are expected in the twin-experiment setup, where the truth and model underlying the assimilation have similar process representations and parameterizations. 

Overall, the results demonstrate that continental MAGIC TWSA assimilation mainly affects soil moisture timing and downward propagation when a higher temporal resolution of the MAGIC observational input is provided and the choice for an univariate or multivariate MAGIC assimilation has only a slight influence.

\subsection{Regional-scale assimilation over European catchments}
\subsubsection{Review and limitations of existing regional GRACE/-FO assimilation frameworks focusing on carbon cycle}
The impact of regional GRACE/-FO TWSA assimilation on different water-cycle variables has been studied extensively in addition to the global and continental systems ascribed before. One of many examples is \citet{tangdamrongsub_data_2015}, where they assimilated TWSA into the wflow\_hbv model over the Rhine River basin and found reductions in groundwater storage RMSE of up to 35\% relative to the OL simulation. They report that the largest improvements in RMSE occurred when the model was driven by coarser global forcing and with uncalibrated parameters, indicating the value of TWSA DA in data sparse regions. For the Murray-Darling Basin, \citet{schumacher2018improving} combined parameter calibration with GRACE assimilation in WaterGAP and showed that the assimilation improved the representation of the Millennium Drought, increasing correlations with in-situ well observations to up to 0.82. 

However, GRACE/-FO TWSA assimilation implications for carbon related variables such as gross primary productivity (GPP) and leaf area index (LAI) remain comparatively unexplored. So far most regional but also continental and global assimilation frameworks assessed utilize models that lack the structure needed to simulate the carbon cycle at all. For instance, CLSM, one of the most widely employed models in the field of TWSA DA \citep{zaitchik_assimilation_2008,girotto_assimilation_2016,li_assimilation_2012}, relies on prescribed vegetation phenology leaving no internal carbon state for a TWSA update to influence. Models that have an available carbon-vegetation option are generally run with that setting disabled. 

One of a few DA systems whose model integrates vegetation is the system developed by \citet{Nie2019Irrigation}. They assimilated GRACE TWSA into the Noah-MP model that uses MODIS derived vegetation greenness to drive an irrigation scheme instead of the model's dynamic-leaf option. CLM3.5 used by \cite{su_multisensor_2010} and in the continental comparison in Section 3.2 includes an optional prognostic carbon-nitrogen configuration (CLM-CN) but has historically been run with satellite prescribed climatology. Lastly, over East Asia, \citet{chi2024clmbgc} assimilated GRACE/-FO TWSA into CLM5 with BGC mode, i.g. carbon cycle variables such as GPP and LAI are explicitly simulated, yet their evaluation covers only TWS anomalies, soil moisture and evapotranspiration. 

Carbon and water cycle improvements from satellite based assimilation have been demonstrated by directly assimilating vegetation-related observations. \citet{kumar2019lai} assimilated LAI into Noah-MP over the continental United States, reducing domain averaged GPP and Net Ecosystem Exchange (NEE) RMSE by 28\% and 5\%, respectively, with larger gains over agricultural areas. They also noted a small positive impact on TWS anomaly correlation (0.45 to 0.48). \citet{Heyvaert2024smlai} found that LAI assimilation into Noah-MP over Europe reduced GPP RMSD by 0.024 gC/m2 day, while soil moisture assimilation alone slightly degraded GPP and NEE skill. Jointly assimilating both did not exceed LAI only carbon cycle skill, but managed to retain most of it along the soil moisture benefits of the assimilation of soil moisture.

We expect TWSA DA to improve the representation of the carbon cycle, since photosynthesis and transpiration draw on water throughout the root zone, which in eCLM extends to depths of up to 2 meters depending on plant functional type \citep{lawrence_community_2019}. While surface soil moisture DA directly constrains only the upper few centimeters, TWSA integrates water storage over the full column and its assimilation modifies root-zone water availability, thereby GPP and LAI estimates. How much this change in root-zone soil moisture can impact the carbon cycle is highly uncertain, for instance the representation of soil moisture stress on photosynthesis alone accounts to 40-80\% of inter-model spread in simulated GPP \citep{trugman2018}. 

A limitation of current gravity missions providing TWSA observations is their monthly temporal resolution. Vegetation responds to water stress on time scales of days to weeks, for instance, during dry spells or drought onset during the growing season. With monthly GRACE/-FO updates, the simulated vegetation state runs on an uncorrected soil water state for up to a month. The 5-daily MAGIC TWSA product could correct the root-zone state closer to the time at which water stress affects photosynthesis. Thus, we investigate the benefit of a higher temporal resolution TWSA product that a future MAGIC mission could provide for a regional DA system and its effects on the carbon cycle variables GPP and LAI. 

\subsubsection{Impact of MAGIC assimilation frequency on TWSA and carbon variables} 
\begin{figure}[h!]
\begin{center}
\includegraphics[width=18cm,trim= 0cm 0cm 0cm 0cm, clip]{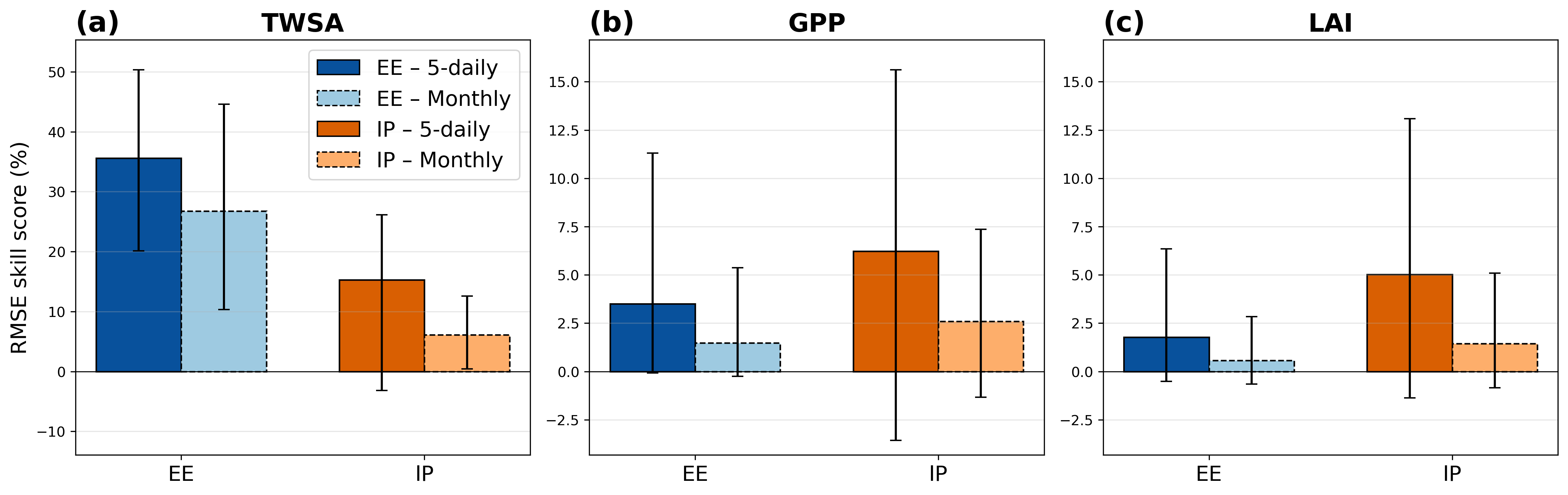}
\end{center}
\caption{Median RMSE skill scores (\%) over the open loop baseline for TWSA (a), GPP (b) and LAI (c), comparing regional monthly and 5-daily MAGIC eCLM assimilation across Eastern Europe (EE) and the Iberian Peninsula (IP). Bar heights give the median RMSE over all 3\,km grid cells of a region; error bars span the inter-quartile range (25th to 75th percentile).}\label{fig:carbon_rmse}
\end{figure}

Assimilating MAGIC TWSA simulations into the regional eCLM model improves the TWSA RMSE towards the simulated truth across both study area, the Iberian Peninsula (IP) and Eastern Europe (EE). The largest improvements of MAGIC DA compared to OL are obtained with the 5-daily assimilation (Fig.~\ref{fig:carbon_rmse}, left), which agrees with the results from the continental MAGIC DA systems. Monthly DA increases median RMSE skill scores (over all 3\,km grid cells of a region) over the open loop by 26.8\% in EE and by a more modest 6.1\% over IP, with 5-daily DA adding a further 9 percentage points in both regions. This RMSE improvement also extends to the representation of the carbon cycle variable GPP and LAI (Fig.~\ref{fig:carbon_rmse}, middle and right). This is consistent with the indirect pathway by which a TWSA update alters soil moisture availability first, before it can influence GPP and LAI. Monthly assimilation alone produces a small, relatively region-independent carbon cycle response, improving median GPP skill by around 1.5\% (EE) and 2.6\% (IP), with lower median LAI skill improvements of 0.6-1.4\%. Five-daily updates during eCLM DA yields additional improvements in both variables GPP and LAI, but its magnitude differs between regions. 

Over the Iberian Peninsula, higher frequency assimilation adds 3.6\% for both GPP and LAI, with a quarter of grid cells showing GPP and LAI improvements exceeding 15.6\% and 13.1\% respectively. These results can be compared to results acquired with direct LAI assimilation, where \citet{fox2022} report a 17.8\% reduction in GPP RMSE, a value that the 5-daily TWSA DA approaches in its most responsive locations. Over Eastern Europe the additional gain is smaller, contrary to the pattern seen in TWSA itself, where Eastern Europe saw the larger improvements in water storage estimates. This would suggest that the strength of the hydrological correction and the strength of its carbon cycle response are not proportional and the water limited Iberian Peninsula is more sensitive to how frequently TWSA updates occur.  

The inter-quartile ranges (IQR) show that these improvements are not spatially uniform. For GPP and LAI, the lower quartile is negative for both assimilation frequencies in both regions, meaning that at least a quarter of grid cells are slightly degraded compared to the OL. With 5-daily updates, the whole IQR shifts towards higher skill over EE, with the lower quartile rising for all three variables. While over IP, the IQR widens in both directions, thus more frequent updates over IP increase the performance in the most responsive grid cells but also lead to degradation in others.

When repeating the analysis after spatially averaging all experiments to 0.5° and 1.5° grids, the median TWSA skill score of the 5-daily MAGIC DA increases from 35.6\% at 3\,km to 47.2\% (0.5°) and 55.9\% (1.5°) over EE and from 15.3\% to 25.1\% and 38.3\% over IP. Simultaneously, the additional gain from 5-daily compared to monthly updates grows to 12\% over EE and 19\% over IP at 1.5° resolution. For LAI the median 5-daily skill increases from 5.0\% to 10.3\% over IP, while remaining lower over EE (1.8\% to 2.4\%). For GPP the median RMSE skill changes only slightly over both regions with higher frequency assimilation to 5.8\% (EE) and 7.2\% (IP).

In interim-summary, we could show that regional 5-daily MAGIC DA would enable not only more frequent updates to TWSA with smaller RMSE than for monthly DA updates, but this improvement would also transplant to the carbon cycle variables GPP and LAI.

\begin{figure}[h!]
\begin{center}
\includegraphics[width=15cm,trim= 0cm 0cm 0cm 0cm, clip]{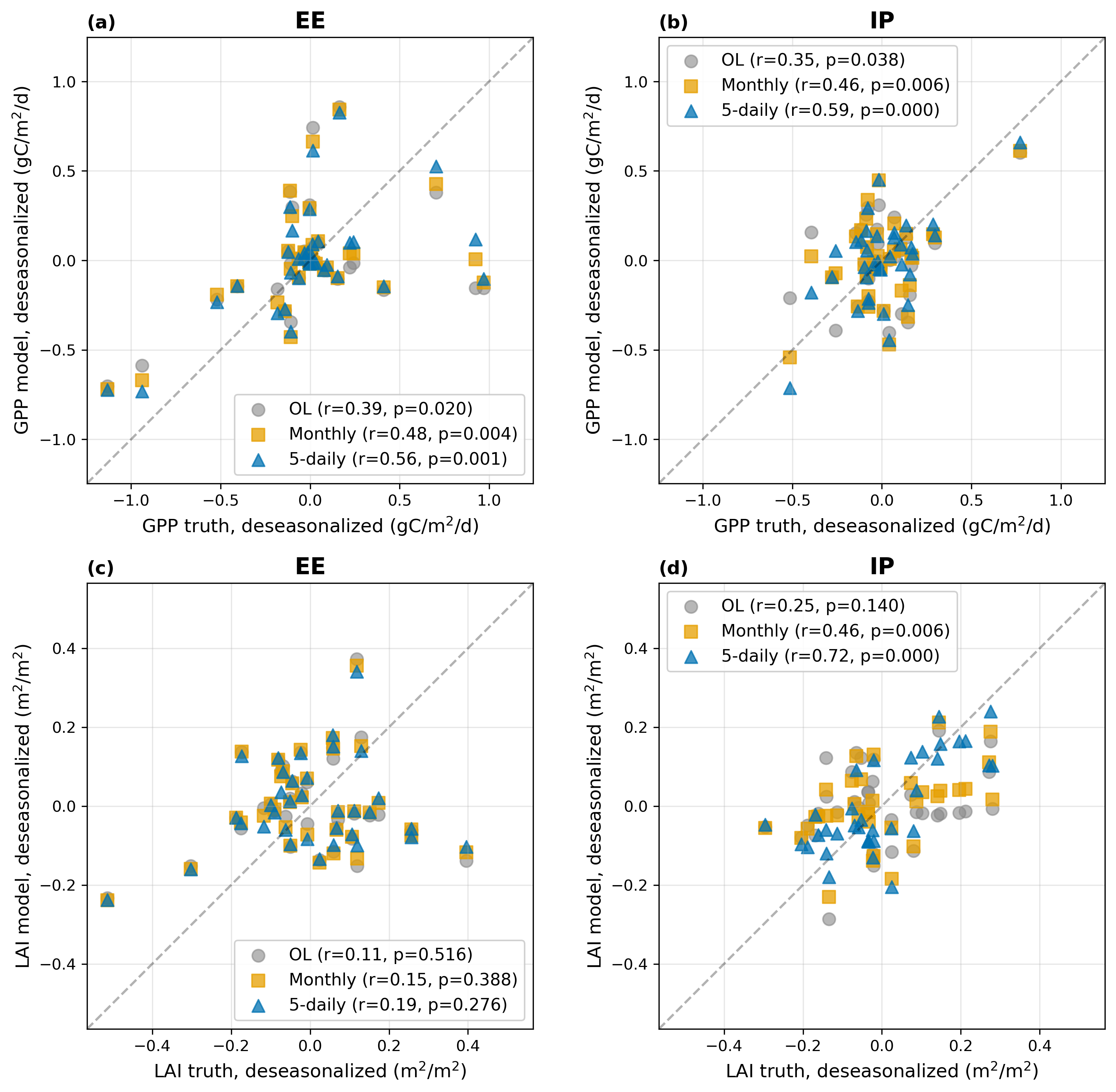}
\end{center}
\caption{Monthly deseasonalized anomalies of region-mean GPP (a,b) and LAI (c,d) over Eastern Europe (a,c) and the Iberian Peninsula (b,d). Model anomalies from the open loop simulation (OL) and regional MAGIC TWSA assimilation with monthly and 5-daily DA updates are plotted against the corresponding truth anomalies. Each point represents one month. The dashed line indicates the 1:1 relationship between model and truth. Pearson correlation coefficients (r) and associated p-values (p) are given in the legend.}\label{fig:carbon_corr}
\end{figure}

The same pattern is visible in the correlations between the deseasonalized simulated and true region-mean GPP and LAI (Fig.~\ref{fig:carbon_corr}, legends). For LAI over the Iberian Peninsula, the OL correlation with the truth is weak and not statistically significant, monthly MAGIC TWSA DA is sufficient to derive a significant correlation towards the truth and 5-daily assimilation strengthens it further to the highest correlation found in this analysis (LAI r=0.72). Over Eastern Europe, by contrast, LAI correlation of the OL towards the truth is similarly weak and non-significant, and neither monthly nor 5-daily DA is able to establish significance. GPP shows a more consistent pattern, the OL correlation towards the truth is already significant in both regions and both assimilation frequencies strengthen it further. Larger additional gain is found for the 5-daily MAGIC TWSA DA again over the Iberian Peninsula. Beyond the correlation, the assimilation also corrects the amplitude of the monthly anomalies over IP (Fig.~\ref{fig:carbon_corr} d), while the OL underestimates positive LAI anomalies, the 5-daily DA moves them closer to the 1:1 line. Over EE, in contrast, the LAI anomalies are nearly identical in all three experiments, this means that the TWSA update only minimally affects the representation of deseasonalized LAI in that region.

These results suggest that a higher frequency TWSA product of the kind planned for the future MAGIC mission, contains valuable additional information beyond what a monthly product provides, and that this information translates into a significant impact on the carbon cycle. This benefit appears to be region dependent: The additional gain from the 5-daily MAGIC assimilation is substantial over the water limited Iberian Peninsula, where vegetation productivity appears more tightly coupled to soil moisture variability. In contrast to the more energy limited Eastern European domain, the same increase in assimilation frequency yields a smaller, and for improvements in LAI correlation, statistically insignificant results. This indicates that TWSA assimilation does not uniformly enhance carbon cycle skill, but its value may be realized most strongly where vegetation is more water-limited.

The regional contrast in LAI and GPP raises the question of how much of this pattern reflects the models representation of water-carbon coupling rather than the response of real ecosystems. In the synthetic setup, the truth and DA experiments share the same parameterizations governing how changes in root-zone water availability affect GPP and LAI. The experiment thus assesses how improvements in the water state estimate impact the simulated carbon cycle, but the response to these improvements is not tested independently. The larger gains over the Iberian Peninsula suggest that eCLM simulates a stronger sensitivity of vegetation productivity to water availability. Although eCLM has been shown to enhance the representation of GPP compared to earlier model generations \citep{lawrence_community_2019}, regional biases remain and might affect how TWSA updates influence estimates of carbon cycle variables.
As with the experiments discussed above, the magnitude of these effects depends on the specific assimilation framework used here and should not be interpreted as a direct prediction of a future MAGIC missions performance. However, the results suggest that a future 5-daily TWSA product could deliver the largest benefit for carbon cycle from regional MAGIC assimilation in water-limited regions.

\section{Discussion}
Overall, our results highlight the potential of using longer time series and more frequent TWSA observations from MAGIC in DA systems to improve the characterization of TWSA and the representation of related storage components, fluxes, and land-surface variables. Extending the current GRACE/-FO observation record with the MAGIC period reduces uncertainties in trends of TWSA and individual storage components such as groundwater from assimilation, allowing smaller trends that are currently difficult to distinguish from observational uncertainty to be detected. This also provides a more robust basis for identifying and classifying drought events, which could support the assessment of drought impacts and management responses. The results further show that more frequent TWSA observations can affect how rapidly DA information is transferred to other components of the land system, with larger DA impacts for 5-daily than for monthly updates during MAGIC DA. Five-daily MAGIC DA leads to faster responses in TWSA, runoff and soil moisture and to larger responses in GPP and LAI than monthly DA. The response of these variables also depends on the model parameterization and hydrological regime, as illustrated by the stronger runoff response in CLM3.5 DA than in eCLM DA and the stronger GPP and LAI responses in water-limited than in energy-limited regions. The multivariate experiments show slight differences in the propagation of droughts into deeper soil layers, which requires further investigation to assess the physical plausibility of vertically resolved DA responses in both univariate and multivariate assimilation.

Our results highlight the potential of including MAGIC in global, continental and regional DA applications, however, they are based on assumptions and simulations and therefore this study is not free of limitations. 

First, the simulation setup within the ESA SING study had to strike a balance between numerical complexity and efficiency. The observation geometry for GRACE-C and NGGM satellites assumes some degree of orbital control which, at least according to the current baseline for NGGM \citep{daras2026nggm}, indeed corresponds to a monthly repeat with 5-day sub-cycle patterns. If the real GRACE-C orbit will be uncontrolled (such as for GRACE and GRACE-FO), there may be months with degraded TWSA data quality due to deep repeat patterns, and this is not represented here. The simulation setup for the satellite scenarios utilizes assumptions on current background model errors which are notoriously difficult to quantify, but we can hope that some progress will be still made in the modeling community such that we are on the conservative side here. Also, the simulation makes use of simplified instrument error models and a simplified retrieval method; however the simulated GRACE-C errors appear quite realistic for a GRACE-FO follower mission.

The SING project seeks to realistically simulate uncertainties that arise from imperfect tidal and non-tidal background modeling via correction model ensembles, and it is impossible to represent the resulting errors in spherical harmonic domain by simple approaches such as scaling the formal variance-covariance matrix. Error samples are consequently provided as an ensemble which can be used to estimate the empirical variance covariance matrix for the GRACE-C-like and MAGIC scenarios. In contrast, real GRACE/-FO data assimilation systems currently work either with constant error assumptions, e.g. 3 cm for each TWSA grid cell, or they use available formal errors derived from the normal equation system of the gravity fields (either variances only, or variances and correlations). Research on estimating empirical variance-covariance matrices from error samples is a quickly developing field, which we believe will become more important for MAGIC error assessments, but so far there is no uniform consensus in the DA community on this. 

It is important to understand that in our OSSE simulations, we had to make assumptions about structural and forcing-data related random and systematic uncertainties in the WaterGAP and CLM models and the truth. If these uncertainties are underestimated, simulated data assimilation will lead to unrealistically low improvements, while if they are over-represented, data assimilation suggests too optimistic improvements when compared to reality. Since hydrological models are under continuous developement, with improved and new process representations (e.g. coupling 3D groundwater modeling), indeed a closer agreement between real MAGIC observations and simulations might be achieved within a decade as compared to our simulations. On the other hand, we can expect that atmospheric and ocean background models will improve at similar pace, with the gravity solutions also improving. It is impossible to map these potential improvements in a limited simulations study here.

At the same time, we also did not attempt to overly tune the level-3 data postprocessing and the DA methods compared to the GRACE/-FO application. In the real world this will happen certainly, at least in the research context. For example, we are convinced that with an improved spatial resolution of the MAGIC data, the MAGIC-derived TWSA maps can be ingested on a significantly finer input grid \citep{kusche2025benefit} as compared to the global 4° or continental and regional 2° resolution as currently used in our systems.

Despite these challenges, we expect that in future the improvements in spatial resolution and temporally extended time series of TWSA from MAGIC assimilation allows a more precise estimate of the essential climate variable TWS, for example, for its use in water balance equation. In addition, TWSA from MAGIC assimilation could also be integrated in scientific applications such as forecasting water storage with machine learning \citep{li_long-term_2021}, correcting GNSS for hydrological loading \citep{springer_evidence_2019} or validating climate models \citep{jensen_observations_2024}. 

Since the assimilation allows vertical disaggregation of TWSA into storages such as groundwater, surface water, and root-zone soil moisture etc., it will also provide a temporal extension and better estimate of these storages. Providing root-zone soil moisture from MAGIC assimilation is promising for agricultural water management services as plants derive their water from the root zone and current soil moisture satellite products are generally limited to the first few centimeters of the soil. For example, the Copernicus Land Monitoring Service \citep{copernicus} could integrate a root zone product. TWSA, root-zone soil moisture, groundwater, runoff and carbon variables such as LAI estimates from assimilation could enhance operational services especially when low-latency MAGIC products will be integrated in DA, for example, the monitoring of agricultural and hydrological drought as in the European and Global Drought Observatories \citep{edo_gdo} or flood monitoring in the Global Flood Awareness System \citep{glofas}. 

Many studies nowadays derive groundwater storage change via a residual approach, where variations in other storages such as in soil moisture, snow, etc. are removed independently \citep[and potentially inconsistent due to incompatible model structures, e.g.][]{getirana2025} from TWSA maps. In contrast, within a DA approach, the estimate for groundwater storage change can be viewed as derived from TWSA within a residual approach where all other storage anomalies are derived from a consistent processing and model structure. Furthermore, groundwater from GRACE/-FO assimilation was already integrated in groundwater risk assessments \citep{wwf}, the importance of assimilation products has already been acknowledged and the continuation of such products with MAGIC assimilation would be ensured. Thus, enhanced water storages from MAGIC assimilation will not only ensure the continuation of existing scientific applications but also enable new applications and bring forward the use of it in operational services due to the improved spatial resolution, uncertainty, reduced latency, and temporal extension.


\renewcommand\thefigure{\thesection.\arabic{figure}}    
\appendix
\setcounter{figure}{0}
\section*{Statements and declarations}   


\subsection*{Author Contribution Statement.}
H.G. developed the paper structure and organized the writing process. All authors contributed to conceiving the experiments, H.G., A.S., A.N., and Y.E. conducted the experiments. All authors contributed to writing sections to derive a first full draft of the manuscript and all authors reviewed the manuscript.

\subsection*{Conflict of Interest Statement.} The authors declare that the research was conducted in the absence of any commercial or financial relationships that could be construed as a potential conflict of interest.


\subsection*{Data Availability Statement.}
The level-2 ESA SING variance-covariance matrices \citep{schlaak2026} for the GRACE-C-like and MAGIC gravity mission simulation scenarios are publicly available through the ISDC data portal at \url{https://isdc-data.gfz.de/future_missions/}. The real-world global WaterGAP assimilation known as GLWS4 \citep{gerdener2026gagp} is publicly available through PANGAEA  \url{https://doi.org/10.1594/PANGAEA.993748}. The simulations with CLM3.5 and eCLM have been carried out within the TSMP–PDAF framework, available at https://github.com/HPSCTerrSys/.

\subsection*{Acknowledgments and Funding.}
H.G. and J.K acknowledge funding from the ESA SING project (https://www.esa-sing.org/), funded by the European Space Agency under the "NGGM and MAGIC Science and Applications Impact Study" (ESA Contract No. 4000145265/24/NL/SC). A.N. acknowledges funding from the European Space Agency (OSIP grant no. 4000147719/2025/NL/GLC). A.S., A.N., Y.E., and J.K. acknowledge funding from the Deutsche Forschungsgemeinschaft (DFG, German Research Foundation) for SFB 1502/1–2022 (Projektnummer: 450058266). We acknowledge Marius Schlaak and Roland Pail for providing the MAGIC and GRACE-C-like simulations in the ESA SING project. We thank Christian Poppe Ter\'an for providing the spun-up restart file for the eCLM-BGC simulation.

\bibliography{references.bib} 
\bibliographystyle{apalike-ejor}

\end{document}